\documentclass{aa}  
\usepackage{graphicx}
\usepackage{url}
\usepackage{natbib}
\usepackage{txfonts}
\usepackage[normalem]{ulem}
\usepackage{longtable}
\usepackage{color}
\usepackage{amssymb,xcolor}
\begin{document} 

   \title{Optical/X-ray correlations during the V404~Cyg June 2015 outburst}
%       \subtitle{Searching for time lags between the optical and X-ray emission}

\titlerunning{Optical/X-ray correlations during the V404~Cyg June 2015 outburst}
   \author{J. Alfonso-Garz\'on
          \inst{1}
          \and C. S\'{a}nchez-Fern\'{a}ndez \inst{2} \and P.A. Charles \inst{3} \and J.M. Mas-Hesse \inst{1}  \and   P. Gandhi \inst{3}  \and \\ M. Kimura \inst{5} \and A. Domingo \inst{1} \and J. Rodriguez \inst{6} \and J. Chenevez \inst{7}
          }

   \institute{Centro de Astrobiolog\'{\i}a -- Departamento de Astrof\'{\i}sica
              (CSIC-INTA), Camino Bajo del Castillo s/n, E-28692 Villanueva de la Ca\~nada, Spain \\ \email{julia@cab.inta-csic.es}
         \and {European Space Astronomy Centre (ESAC), Camino Bajo del Castillo s/n, E-28692 Villanueva de la Ca\~nada, Spain} \and{Department of Physics and Astronomy, University of Southampton, Southampton SO17 1BJ, UK} \and{Department of Astronomy, Graduate School of Science, Kyoto University, Oiwakecho, Kitashirakawa, Sakyo-ku, Kyoto 606-8502, Japan} \and{Laboratoire AIM (CEA/IRFU - CNRS/INSU - Université Paris Diderot), CEA DSM/IRFU/SAp, 91191, Gif-sur-Yvette, France} \and{DTU Space—National Space Institute, Technical University of Denmark, Elektrovej 327-328, DK-2800 Lyngby, Denmark}\\
             }

   \date{Accepted: 7 October 2018.}

% \abstract{}{}{}{}{} 
% 5 {} token are mandatory
 
  \abstract
   {We present a multiwavelength analysis of the simultaneous optical and X-ray light curves of the microquasar V404~Cyg during the June 2015 outburst.}
   {We have performed a comprehensive analysis of all the \textit{INTEGRAL}/IBIS, JEM--X, and OMC observations during the brightest epoch of the outburst, along with complementary \textit{NuSTAR}, AAVSO, and VSNET data, to examine the timing relationship between the simultaneous optical and X-ray light curves, in order to understand the emission mechanisms and physical locations.}
   {We have identified all optical flares which have simultaneous X-ray observations, and performed cross-correlation analysis to estimate the  time delays between the optical and soft and hard X-ray emission. We have also compared the evolution of the optical and X-ray emission with the hardness-ratios.}
   {We have identified several types of behaviour during the outburst. On many occasions, the optical flares occur simultaneously with X-ray flares, but at other times positive and negative time delays between the optical and X-ray emission are measured.}
   {We conclude that the observed optical variability is driven by different physical mechanisms, including reprocessing of X-rays in the accretion disc and/or the companion star, interaction of the jet ejections with surrounding material or with previously ejected blobs, and synchrotron emission from the jet.}


   \maketitle
%
%-------------------------------------------------------------------

\section{Introduction}\label{sec:int}

Multiwavelength observations of black hole X-ray binaries (BHXBs) in outburst are crucial to understand the physical mechanisms at work in the different regions of the accretion flow. There is a general consensus that the X-rays are produced in the vicinity of the compact object and the radio emission is associated with the jet. However, the origin of the optical emission in BHXBs can be generated by a variety of mechanisms, taking place in different regions of the system. These may include: reprocessing of X-rays in the outer regions of the accretion disc and in the optical companion, self-synchrotron emission in the inner hot flow, and synchrotron emission at the base of the jet \citep{lewin2006,hynes2010, gandhi2016}. 

Performing cross-correlation analysis of the optical and X-ray light curves is an essential tool to characterize the physical mechanisms driving the observed optical fluxes, and ideally select amongst competing physical models, as different emission regions would lead to different observed time lags. It is commonly accepted that in many cases the optical emission is X-ray driven, mainly as reprocessing in the accretion disc and in the optical companion \citep{vanparadijs1994,dubus1999}. In this scenario, light travel time delays between the X-rays and the optical emission are expected. Such time lags have been observed e.g. in Sco~X-1 \citep{munoz-darias2007,hynes2016}. However, there are other mechanisms (e.g. synchrotron emission from the jet) which can lead to sub-seconds lags and produce more complex patterns in the optical/X-ray correlation functions (as has been observed for XTE~J1118+480 \citealt{kanbach2001,hynes2003}, and GX~339-4 \citealt{markoff2001, gandhi2008}).

V404~Cygni (hereafter V404~Cyg) is a transient low mass X-ray binary (LMXB) consisting of a black hole (BH) accreting mass from a low-mass optical companion. The distance to the system is 2.39$\pm$0.14\,kpc \citep{miller-jones2009}. The orbital period is 6.47\,d and the mass function is $f(M)$=6.08$\pm$0.04\,M$_\sun$  \citep{casares1994}. The inclination of the orbit is in the range 56--67$^{\circ}$ \citep{shahbaz1994,sanwal1996,khargharia2010}. The donor star is a K0~III--K2~IV \citep{khargharia2010,gonzalezhernandez2011}, and the mass of the black hole is M$_{BH}$ = 9--12\,M$_{\sun}$.

V404~Cyg was first detected in the optical and identified as a nova-like system, during its outbursts in 1938 \citep{wachman1948}, 1956, and probably 1979 \citep{richter1989}. The X-ray source GS~2023+338, discovered in X-rays by the Ginga satellite during an X-ray outburst in May 1989 \citep{makino1989-a} was identified as the old Nova Cygni 1938 (V404~Cygni, \citealt{wagner1989}). On 15 June 2015 18:32~UTC (MJD~57188.772), after 26 years of quiescence, the Burst Alert Telescope (BAT, \citealt{barthelmy2005}) onboard the \textit{Swift} satellite \citep{gehrels2004} detected renewed X-ray activity from V404~Cyg \citep{barthelmy2015}. This outburst reached its peak on 26 June 2015 and then rapidly decayed, returning to quiescence in mid-August \citep{sivakoff2015}. The outburst was intensively observed by most of the available  space telescopes and ground-based facilities worldwide. On 23 December 2015 (MJD~57379), the Swift/BAT instrument detected a new outburst from V404~Cyg \citep{barthelmy2005}. The 2015 December outburst was shorter and fainter than the June 2015 outburst \citep{kajava2018}.

Several studies comparing the optical and X-ray emission from these two outbursts have been performed during the last years, finding no lags or lags from fractions of seconds to several seconds. \citet{maitra2017} did not find lags between their optical observations and the X-ray light curves during the June 2015 outburst and proposed the origin of the optical emission to be a compact region most likely originating in a jet outflow. \citet{gandhi2017} measured optical lags of 0.1\,s in two epochs of their June 2015 observations, which they attributed to the elevation of the optical jet base of the system. \citet{kimura2017,kimura2018} found evidence of lags of 22.5 and 34.8\,s between the optical and X-ray emission for two epochs in their observations during the December 2015 outburst. \citet{gandhi2015} measured lags of several seconds between the optical and X-rays during their observations from 21 June 2015. In both works, the authors attributed these optical lags of several seconds to reprocessing of the X-ray emission in the accretion disc. \citet{rodriguez2015} studied the available \textit{INTEGRAL} observations in the period from 20--25 June 2015, and found optical lags of several minutes, pointing to a jet origin. 

In this paper, we analyse all the available \textit{INTEGRAL} data during the brightest period of the June 2015 outburst, i.e. from 18 June 2015, to the fast decay of the optical and X-ray emission in 27 June 2015, including also \textit{NuSTAR} observations performed from 25--26 June 2015, and some complementary public optical observations.

%--------------------------------------------------------------------
\section{Observations}\label{sec:obs}

The INTErnational Gamma-Ray Astrophysics Laboratory (\textit{INTEGRAL}, \citealt{winkler2003})  observed V404~Cyg in the period MJD~57191.5 to 57216. We analyse here the data obtained during the interval MJD~57191.75 to 57200.25, when intense flaring activity was detected from the system. We study the optical light curves in the $V$--band provided by the Optical Monitoring Camera (OMC, \citealt{mas-hesse2003}), and the X-ray light curves provided by the Joint European X-Ray Monitor (JEM-X, \citealt{lund2003}; 3--10\,keV), and by the Imager on Board the \textit{INTEGRAL} Satellite (IBIS, \citealt{ubertini2003}; 20--80\,keV and 80--200\,keV).

The \textit{INTEGRAL} data reduction was performed with the Off-line Science Analysis (OSA; \citealt{courvoisier2003}) software version 10.2, using the latest calibration files. The data were processed following standard reduction procedures.

We also added to our analysis the X-ray light curves provided by the Nuclear Spectroscopic Telescope Array (\textit{NuSTAR}, \citealt{harrison2013}) in the 3--10 and 10--79\, keV bands. These observations were performed during MJD~57197.94--57199.05 \citep{walton2017}.

The \textit{INTEGRAL}/JEM-X and IBIS light curves have been binned to provide a time resolution of 5\,s. We use a time resolution of 1\,s for the \textit{NuSTAR} light curves.

We complement the OMC optical monitoring with public observations provided by the American Association of Variable Star Observers (AAVSO, \citealt{kafka2018}) and with data from the Variable Star Network (VSNET) Collaboration (see \citealt{kimura2016} for a description of this dataset). We converted the reference times provided by the observers in Coordinated Universal Time (UTC) to Barycentric Dynamical Time (TDB), in order to correctly compare the ground-based data with the \textit{INTEGRAL} and \textit{NuSTAR} observations. In some cases we have applied small flux corrections for some observers' data sets, always taking as reference the OMC fluxes. All the optical observations have been corrected for interstellar extinction assuming a value of A$_{V}$=4\,mag \citep{casares1993}.

The optical monitoring mean cadences vary between 3\,s and 260\,s along the outburst.

\section{Data analysis}

We show the V404~Cyg optical and X-ray light curves from MJD~57191.75 to MJD~57200.25 analysed in this work in Figs. \ref{fig:LCcomplete} and \ref{fig:LCcomplete2}. In this section we describe the properties of these light curves and the analysis we have performed.

\subsection{Identification of optical flares}\label{sec:optflares}

\begin{figure*}[tp]
   \includegraphics[width=\textwidth]{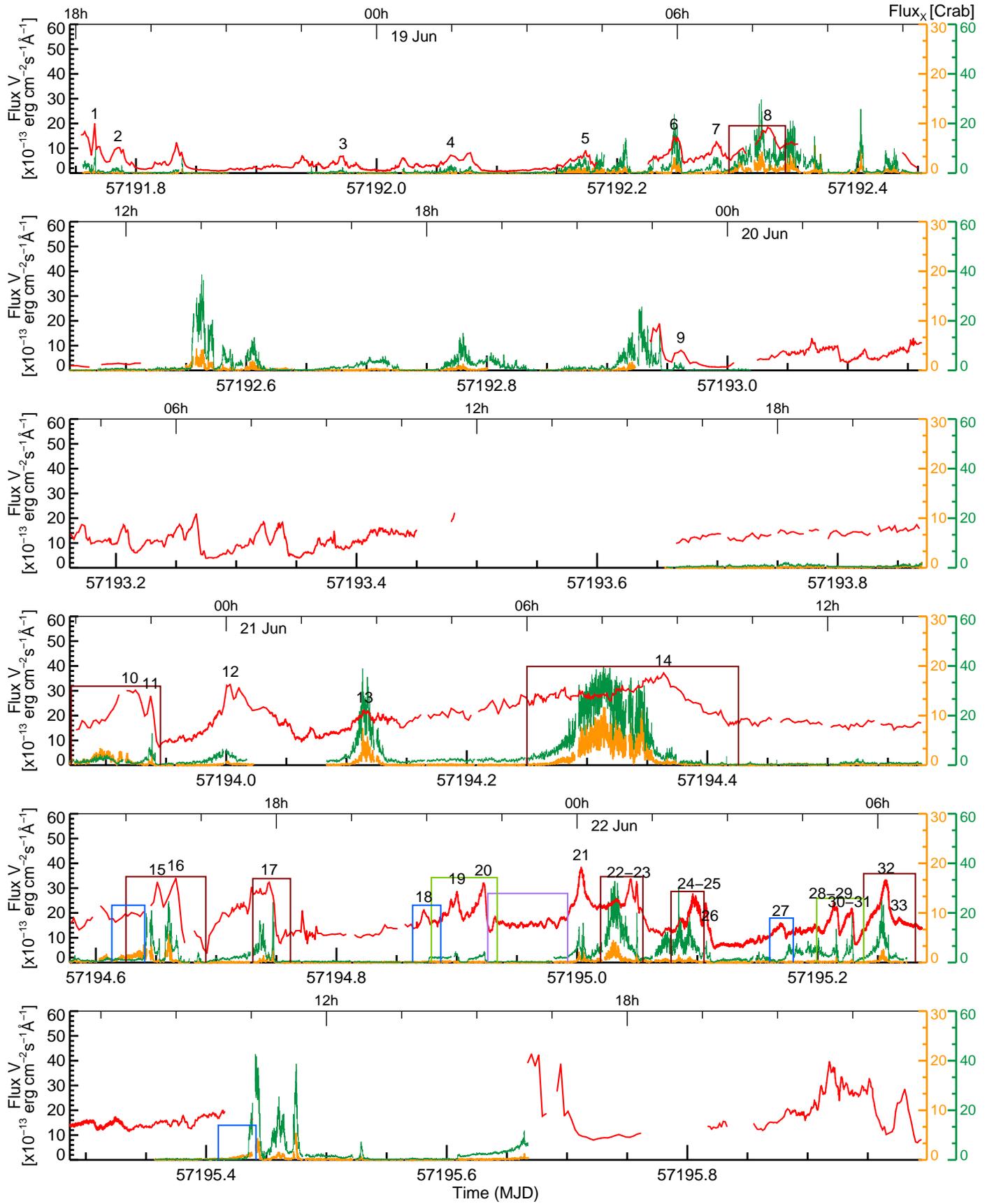}
\caption[Light curves of V404~Cyg]{Optical and X-ray light curves of V404~Cyg during the June 2015 outburst from MJD~57191.75 to MJD~57196.00. The optical observations are plotted in red, the soft X-ray emission in the 3--10\,keV band is plotted in orange, and the hard X-ray emission in the 20--80\,keV band for \textit{INTEGRAL}/IBIS and in the 10--79 \,keV band for the \textit{NuSTAR} observations are plotted both in green. The numbers in black refer to the optical flare identifications. Brown and blue boxes mark flares with positive and negative lags respectively. Light green boxes represent double symmetric optical flares with simultaneous X-ray emission. Purple boxes represent epochs where `heartbeat-type' oscillations are observed (see text).}\label{fig:LCcomplete}
\end{figure*}

\begin{figure*}[htp]
   \includegraphics[width=\textwidth]{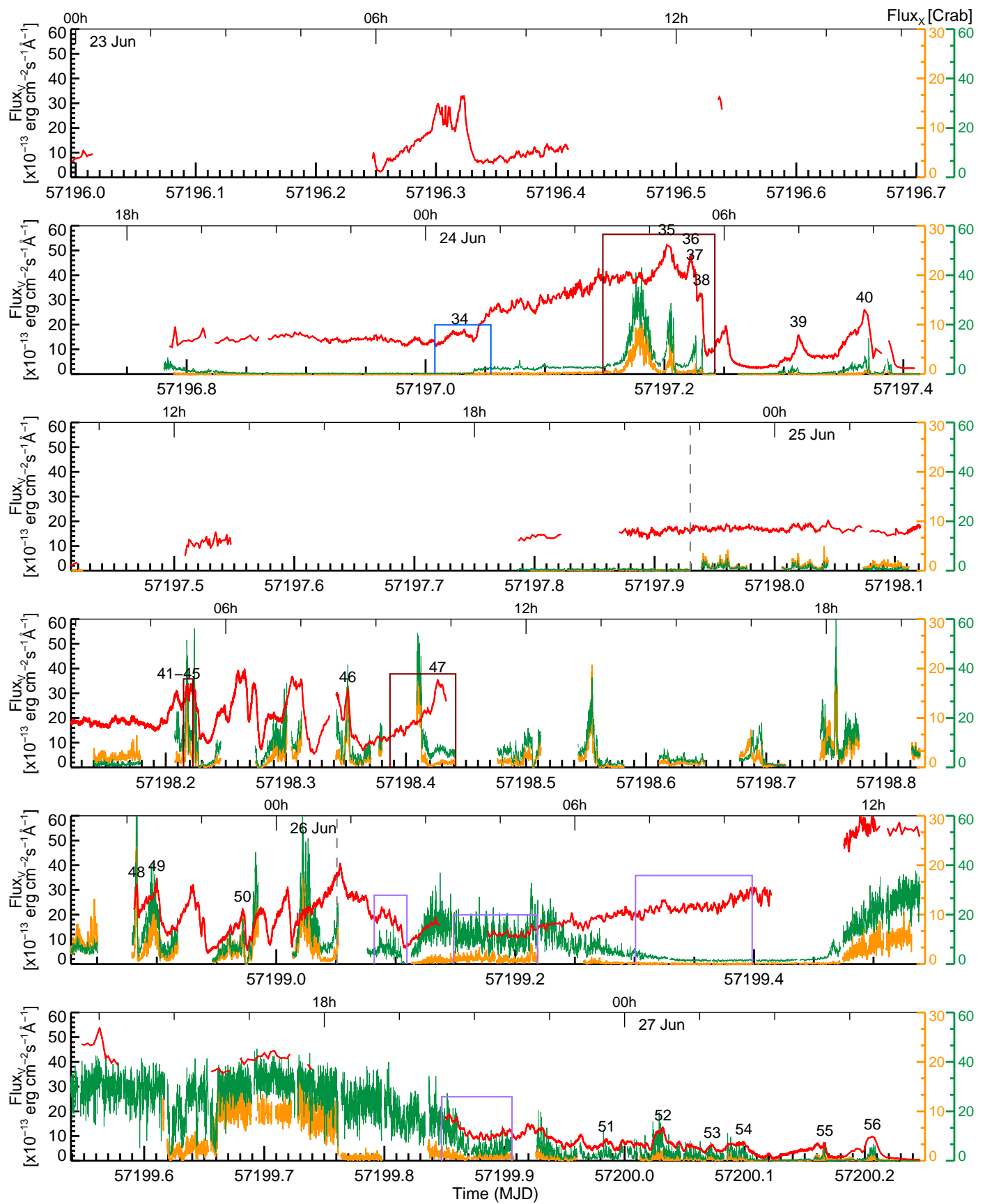}
\caption[Light curves of V404~Cyg]{Same as Fig.~\ref{fig:LCcomplete} but from MJD~57196.75 to MJD 57200.25. Gray dashed lines mark the initial and end times of the \textit{NuSTAR} observations.}\label{fig:LCcomplete2}
\end{figure*}

We have identified all the optical flares which have simultaneous X-ray observations. These flares are flagged with identifying numbers in Figs.~\ref{fig:LCcomplete} and \ref{fig:LCcomplete2}. The main properties of these optical flares are summarized in Table~\ref{tab:flares}, where for each flare we provide, the time of the peak, the flare rise and decay times, the total flare duration, the peak flux, the flux increment during the flare rise ($\Delta$F$_{rise}$) and the flux decrement during the decay ($\Delta$F$_{decay}$). 

\begin{figure}[h]
  \includegraphics[width=8cm]{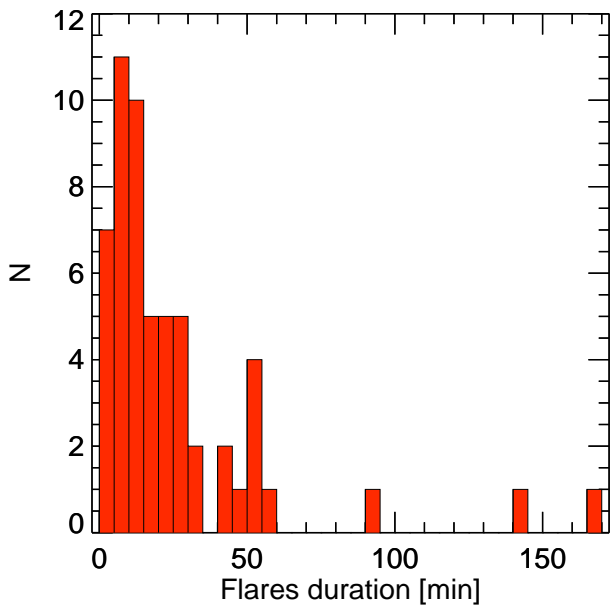}  \\       
 \caption{Histogram of the duration of the optical flares during the June 2015 outburst of V404~Cyg. Bins of 5\,min have been used.}    
\label{fig:histo_fdurat}
\end{figure}

The flare start and end times have been identified as the times when a change in the slope of the optical light curve is detected. The time of the peak has been defined as the time when the optical flux is maximum in each flare. Flare durations, defined as the difference between the flare start and end times, typically range from 3 to 30 min, although some longer flares have been observed (e.g. flares 12 and 14 with durations of 166 and 142\,min respectively). The distribution of the optical flare durations is provided in Fig.~\ref{fig:histo_fdurat}.

\begin{figure}[h]
  \includegraphics[width=8cm]{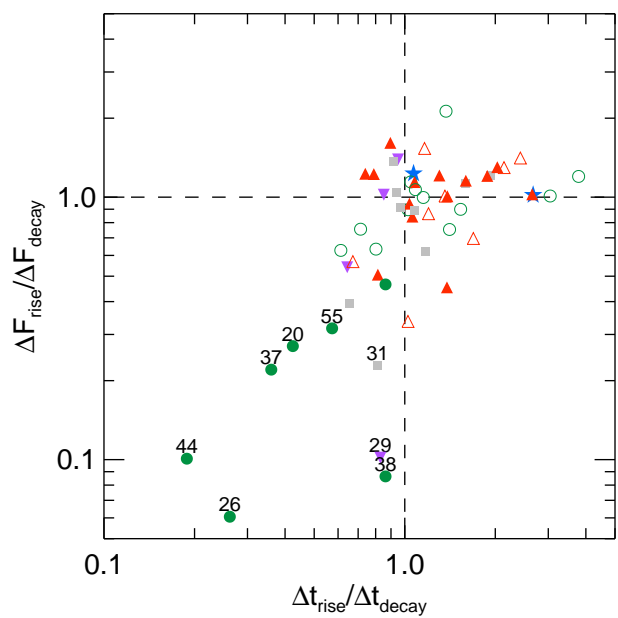}  \\       
 \caption{$\Delta$F$_{rise}$/$\Delta$F$_{decay}$ vs $\Delta$t$_{rise}$/$\Delta$t$_{decay}$ ratios for all the optical flares identified in this work. Dashed lines mark where these properties are symmetrical. Colour coding according to the optical/X-ray lags (see Sect.~\ref{sec:cross}): zero-lagged flares - green points, positive-lagged flares - red triangles, negative-lagged flares - blue stars, frequency dependent lagged flares - inverted purple triangles, uncertain - gray squares. Flare numbers of highly asymmetric flares with longer time decays are labelled.}    
\label{fig:rise_fade}
\end{figure}

In order to study the shape of the flares, we have represented the $\Delta$F$_{rise}$/$\Delta$F$_{decay}$ vs $\Delta$t$_{rise}$/$\Delta$t$_{decay}$ in Fig.~\ref{fig:rise_fade}.  We have used colours to distinguish between the different types of flares, according to the classification we will describe in Sect.~\ref{sec:cross}. The horizontal dashed line corresponds to flares for which the source flux before and after the flare is the same ($\Delta$F$_{rise}$/$\Delta$F$_{decay}$=1). Points above this line correspond to flares for which the optical flux decays to higher fluxes than the flux measured at the flare onset. Points below this line correspond to flares for which the flare decays to lower fluxes than those measured at the flare onset. The vertical dashed line corresponds to flares for which the rise and decay times are equal ($\Delta$t$_{rise}$/$\Delta$t$_{decay}$=1). Points to the left of this line correspond to flares with shorter rise than decay times, whereas points to the right of this line correspond to flares with longer rise than decay times. Most of the flares are symmetric, showing similar $\Delta$F$_{rise}$ and $\Delta$F$_{decay}$, and similar $\Delta$t$_{rise}$ and $\Delta$t$_{decay}$. However, there are some highly asymmetric flares which show longer  time decays, with end fluxes below the pre-flare flux values. \linebreak %\newline

\clearpage
\onecolumn

\clearpage

\clearpage

\begin{table*}[htp]
\caption{Flux and temporal properties of the optical flares identified.}
\label{tab:flares}
\begin{tabular}{r|r|rrr|rrr}
\hline
Flare  & Time$_{peak}$  &$\Delta$t$_{rise}$ & $\Delta$t$_{decay}$ &$\Delta$t$_{total}$ & Flux$_{peak}$ &$\Delta$F$_{rise}$ &$\Delta$F$_{decay}$ \\
\multicolumn{1}{c|}{\#} & \multicolumn{1}{r|}{(MJD-57190)} & \multicolumn{3}{c|}{[min]}& \multicolumn{3}{c}{[$\times$10$^{-13}$erg\,cm$^{-2}$\,s$^{-1}$\,$\AA^{-1}$]} \\

      \hline
\setlength{\tabcolsep}{2pt}
       1 & 1.7657 $\pm$ 0.0005 & 4.0$\pm$0.8 & 2.1$\pm$0.8 & 6.1$\pm$0.8 & 18.9$\pm$0.1 & 12.0$\pm$0.1 & 9.9$\pm$0.1\\
       2 & 1.7851 $\pm$ 0.0007 & 13.7$\pm$1.0 & 10.1$\pm$1.0 & 23.8$\pm$1.0 & 9.77$\pm$0.08 & 5.53$\pm$0.08 & 5.47$\pm$0.08\\
       3 & 1.9722 $\pm$ 0.0003 & 21.1$\pm$0.5 & 6.9$\pm$0.5 & 28.0$\pm$0.5 & 6.83$\pm$0.03 & 4.77$\pm$0.03 & 4.72$\pm$0.03\\
       4 & 2.0616 $\pm$ 0.0007 & 15.8$\pm$1.0 & 11.5$\pm$1.0 & 27.3$\pm$1.0 & 7.25$\pm$0.04 & 4.61$\pm$0.04 & 2.17$\pm$0.04\\
       5 & 2.1737 $\pm$ 0.0005 & 41.4$\pm$0.7 & 10.9$\pm$0.7 & 52.3$\pm$0.7 & 9.07$\pm$0.05 & 7.52$\pm$0.05 & 6.27$\pm$0.05\\
       6 & 2.2473 $\pm$ 0.0004 & 19.1$\pm$0.5 & 12.4$\pm$0.5 & 31.5$\pm$0.5 & 14.6$\pm$0.1 & 9.6$\pm$0.1 & 10.7$\pm$0.1\\
       7 & 2.2824 $\pm$ 0.0003 & 15.0$\pm$0.5 & 13.0$\pm$0.5 & 28.0$\pm$0.5 & 12.6$\pm$0.1 & 6.8$\pm$0.1 & 6.8$\pm$0.1\\
       8 & 2.325 $\pm$ 0.001 & 27.0$\pm$2 & 14.4$\pm$2 & 41.4$\pm$2 & 18.4$\pm$0.3 & 10.3$\pm$0.3 & 8.6$\pm$0.3\\
       9 & 2.961 $\pm$ 0.001 & 11.3$\pm$2 & 11.0$\pm$2 & 22.3$\pm$2 & 8.1$\pm$0.1 & 4.8$\pm$0.1 & 5.4$\pm$0.1\\
      10 & 3.920 $\pm$ 0.001 & 36.3$\pm$2 & 17.8$\pm$2 & 54.1$\pm$2 & 30.2$\pm$0.5 & 13.3$\pm$0.5 & 10.2$\pm$0.5\\
      11 & 3.937 $\pm$ 0.001 & 6.8$\pm$2 & 10.4$\pm$2 & 17.2$\pm$2 & 28.0$\pm$0.4 & 8.0$\pm$0.4 & 20.4$\pm$0.4\\
      12 & 4.0034 $\pm$ 0.0003 & 80.5$\pm$0.4 & 85.6$\pm$0.4 & 166.1$\pm$0.4 & 32.6$\pm$0.2 & 23.4$\pm$0.2 & 22.5$\pm$0.2\\
      13 & 4.1150 $\pm$ 0.0002 & 49.6$\pm$0.2 & 42.6$\pm$0.2 & 92.2$\pm$0.2 & 22.2$\pm$0.2 & 10.6$\pm$0.2 & 6.9$\pm$0.2\\
      14 & 4.3635 $\pm$ 0.0008 & 64.$\pm$1. & 79.$\pm$1. & 143.$\pm$1. & 37.2$\pm$0.2 & 10.6$\pm$0.2 & 20.9$\pm$0.2\\
      15 & 4.651 $\pm$ 0.001 & 7.6$\pm$2 & 7.4$\pm$2 & 15.0$\pm$2 & 32.5$\pm$0.6 & 9.5$\pm$0.6 & 10.1$\pm$0.6\\
      16 & 4.666 $\pm$ 0.001 & 14.8$\pm$2 & 10.7$\pm$2 & 25.5$\pm$2 & 34.0$\pm$0.5 & 11.6$\pm$0.5 & 25.6$\pm$0.5\\
      17 & 4.7441 $\pm$ 0.0005 & 4.6$\pm$0.8 & 4.4$\pm$0.8 & 9.0$\pm$0.8 & 32.4$\pm$0.3 & 4.5$\pm$0.3 & 5.4$\pm$0.3\\
      18 & 4.8727 $\pm$ 0.0002 & 7.8$\pm$0.2 & 7.3$\pm$0.2 & 15.1$\pm$0.2 & 21.36$\pm$0.07 & 7.04$\pm$0.07 & 5.72$\pm$0.07\\
      19 & 4.90022 $\pm$ 0.00009 & 4.8$\pm$0.1 & 7.5$\pm$0.1 & 12.3$\pm$0.1 & 28.7$\pm$0.1 & 6.3$\pm$0.1 & 11.7$\pm$0.1\\
      20 & 4.92213 $\pm$ 0.00008 & 3.2$\pm$0.1 & 7.5$\pm$0.1 & 10.7$\pm$0.1 & 32.1$\pm$0.1 & 5.3$\pm$0.1 & 19.5$\pm$0.1\\
      21 & 5.00348 $\pm$ 0.00006 & 27.61$\pm$0.09 & 29.07$\pm$0.09 & 56.68$\pm$0.09 & 38.4$\pm$0.2 & 23.2$\pm$0.2 & 16.6$\pm$0.2\\
      22 & 5.0445 $\pm$ 0.0001 & 7.2$\pm$0.2 & 5.5$\pm$0.2 & 12.7$\pm$0.2 & 33.6$\pm$0.4 & 10.7$\pm$0.4 & 8.8$\pm$0.4\\
      23 & 5.0495 $\pm$ 0.0001 & 1.9$\pm$0.2 & 3.1$\pm$0.2 & 5.0$\pm$0.2 & 32.8$\pm$0.4 & 8.0$\pm$0.4 & 12.8$\pm$0.4\\
      24 & 5.09850 $\pm$ 0.00002 & 11.02$\pm$0.03 & 7.95$\pm$0.03 & 18.97$\pm$0.03 & 26.9$\pm$0.4 & 12.3$\pm$0.4 & 12.2$\pm$0.4\\
      25 & 5.10693 $\pm$ 0.00002 & 4.33$\pm$0.03 & 4.01$\pm$0.03 & 8.34$\pm$0.03 & 24.0$\pm$0.3 & 9.3$\pm$0.3 & 10.4$\pm$0.3\\
      26 & 5.11048 $\pm$ 0.00002 & 1.43$\pm$0.03 & 5.45$\pm$0.03 & 6.88$\pm$0.03 & 13.9$\pm$0.2 & 0.5$\pm$0.2 & 7.9$\pm$0.2\\
      27 & 5.16931 $\pm$ 0.00002 & 18.14$\pm$0.03 & 6.79$\pm$0.03 & 24.93$\pm$0.03 & 15.9$\pm$0.2 & 6.4$\pm$0.2 & 6.3$\pm$0.2\\
      28 & 5.21332 $\pm$ 0.00002 & 1.01$\pm$0.03 & 1.19$\pm$0.03 & 2.20$\pm$0.03 & 24.0$\pm$0.2 & 2.6$\pm$0.2 & 2.5$\pm$0.2\\
      29 & 5.21603 $\pm$ 0.00002 & 2.79$\pm$0.03 & 3.37$\pm$0.03 & 6.16$\pm$0.03 & 22.4$\pm$0.2 & 1.1$\pm$0.2 & 11.1$\pm$0.2\\
      30 & 5.22508 $\pm$ 0.00002 & 1.62$\pm$0.03 & 1.55$\pm$0.03 & 3.17$\pm$0.03 & 19.7$\pm$0.2 & 1.5$\pm$0.2 & 1.3$\pm$0.2\\
      31 & 5.22837 $\pm$ 0.00002 & 3.26$\pm$0.03 & 4.01$\pm$0.03 & 7.27$\pm$0.03 & 21.9$\pm$0.2 & 3.3$\pm$0.2 & 14.5$\pm$0.2\\
      32 & 5.25704 $\pm$ 0.00002 & 8.75$\pm$0.03 & 3.30$\pm$0.03 & 12.05$\pm$0.03 & 33.0$\pm$0.3 & 11.2$\pm$0.3 & 11.0$\pm$0.3\\
      33 & 5.26760 $\pm$ 0.00002 & 4.36$\pm$0.03 & 5.90$\pm$0.03 & 10.26$\pm$0.03 & 18.1$\pm$0.2 & 2.8$\pm$0.2 & 2.3$\pm$0.2\\
      34 & 7.0287 $\pm$ 0.0002 & 25.9$\pm$0.3 & 16.3$\pm$0.3 & 42.2$\pm$0.3 & 17.4$\pm$0.2 & 5.0$\pm$0.2 & 4.4$\pm$0.2\\
      35 & 7.2019 $\pm$ 0.0002 & 24.7$\pm$0.3 & 22.9$\pm$0.3 & 47.6$\pm$0.3 & 53.5$\pm$0.8 & 16.0$\pm$0.8 & 14.0$\pm$0.8\\
      36 & 7.22212 $\pm$ 0.00010 & 6.3$\pm$0.1 & 3.9$\pm$0.1 & 10.2$\pm$0.1 & 49.9$\pm$0.4 & 10.3$\pm$0.4 & 8.9$\pm$0.4\\
      37 & 7.22549 $\pm$ 0.00010 & 0.9$\pm$0.1 & 2.6$\pm$0.1 & 3.5$\pm$0.1 & 43.3$\pm$0.3 & 2.8$\pm$0.3 & 12.6$\pm$0.3\\
      38 & 7.23110 $\pm$ 0.00008 & 5.4$\pm$0.1 & 6.2$\pm$0.1 & 11.6$\pm$0.1 & 33.1$\pm$0.2 & 2.2$\pm$0.2 & 25.2$\pm$0.2\\
      39 & 7.3121 $\pm$ 0.0001 & 25.2$\pm$0.1 & 27.5$\pm$0.1 & 52.7$\pm$0.1 & 16.1$\pm$0.2 & 12.9$\pm$0.2 & 9.4$\pm$0.2\\
      40 & 7.368 $\pm$ 0.001 & 5.4$\pm$2 & 5.6$\pm$2 & 11.0$\pm$2 & 26.1$\pm$0.7 & 6.5$\pm$0.7 & 7.1$\pm$0.7\\
      41 & 8.21768 $\pm$ 0.00002 & 2.29$\pm$0.03 & 2.11$\pm$0.03 & 4.40$\pm$0.03 & 33.3$\pm$0.3 & 6.1$\pm$0.3 & 5.7$\pm$0.3\\
      42 & 8.22090 $\pm$ 0.00002 & 2.27$\pm$0.03 & 2.87$\pm$0.03 & 5.14$\pm$0.03 & 33.9$\pm$0.3 & 6.3$\pm$0.3 & 5.1$\pm$0.3\\
      43 & 8.22425 $\pm$ 0.00002 & 2.01$\pm$0.03 & 1.96$\pm$0.03 & 3.97$\pm$0.03 & 31.5$\pm$0.2 & 2.9$\pm$0.2 & 8.6$\pm$0.2\\
      44 & 8.22607 $\pm$ 0.00002 & 0.72$\pm$0.03 & 3.83$\pm$0.03 & 4.55$\pm$0.03 & 24.0$\pm$0.2 & 1.2$\pm$0.2 & 12.3$\pm$0.2\\
      45 & 8.23099 $\pm$ 0.00002 & 3.17$\pm$0.03 & 4.71$\pm$0.03 & 7.88$\pm$0.03 & 13.7$\pm$0.2 & 2.2$\pm$0.2 & 3.8$\pm$0.2\\
      46 & 8.3516 $\pm$ 0.0002 & 4.4$\pm$0.2 & 5.5$\pm$0.2 & 9.9$\pm$0.2 & 31.5$\pm$0.2 & 12.2$\pm$0.2 & 19.3$\pm$0.2\\
      47 & 8.4264 $\pm$ 0.0002 & 8.9$\pm$0.3 & 9.9$\pm$0.3 & 18.8$\pm$0.3 & 35.5$\pm$0.2 & 13.8$\pm$0.2 & 8.6$\pm$0.2\\
      48 & 8.8830 $\pm$ 0.0002 & 3.3$\pm$0.3 & 2.3$\pm$0.3 & 5.6$\pm$0.3 & 32.3$\pm$0.3 & 9.9$\pm$0.3 & 13.2$\pm$0.3\\
      49 & 8.9000 $\pm$ 0.0001 & 21.5$\pm$0.2 & 12.7$\pm$0.2 & 34.2$\pm$0.2 & 34.7$\pm$0.3 & 15.8$\pm$0.3 & 22.8$\pm$0.3\\
      50 & 8.97187 $\pm$ 0.00008 & 4.9$\pm$0.1 & 5.6$\pm$0.1 & 10.5$\pm$0.1 & 22.3$\pm$0.2 & 7.7$\pm$0.2 & 16.6$\pm$0.2\\
      51 & 9.98513 $\pm$ 0.00005 & 11.18$\pm$0.08 & 9.32$\pm$0.08 & 20.50$\pm$0.08 & 8.91$\pm$0.08 & 3.90$\pm$0.08 & 4.51$\pm$0.08\\
      52 & 10.03205 $\pm$ 0.00006 & 15.26$\pm$0.09 & 6.30$\pm$0.09 & 21.56$\pm$0.09 & 13.4$\pm$0.1 & 8.9$\pm$0.1 & 6.3$\pm$0.1\\
      53 & 10.07307 $\pm$ 0.00009 & 8.9$\pm$0.1 & 4.2$\pm$0.1 & 13.1$\pm$0.1 & 6.4$\pm$0.2 & 3.1$\pm$0.2 & 2.4$\pm$0.2\\
      54 & 10.09927 $\pm$ 0.00003 & 27.76$\pm$0.05 & 23.71$\pm$0.05 & 51.47$\pm$0.05 & 7.66$\pm$0.04 & 3.97$\pm$0.04 & 6.42$\pm$0.04\\
      55 & 10.16686 $\pm$ 0.00002 & 2.91$\pm$0.03 & 5.07$\pm$0.03 & 7.98$\pm$0.03 & 7.28$\pm$0.05 & 1.81$\pm$0.05 & 5.72$\pm$0.05\\
      56 & 10.2064 $\pm$ 0.0008 & 12.$\pm$1. & 17.$\pm$1. & 29.$\pm$1. & 9.79$\pm$0.05 & 6.41$\pm$0.05 & 8.49$\pm$0.05\\

\hline 
\end{tabular}
%\end{longtable}
\end{table*}
\clearpage
\twocolumn

\noindent As we will see in Sect.~\ref{sec:cross}, most of these asymmetric flares are flares with simultaneous X-ray and optical emission (green filled points: e.g. flares 20, 26, 37--38, 44, 55).

\begin{figure}[h]
   \includegraphics[width=8cm]{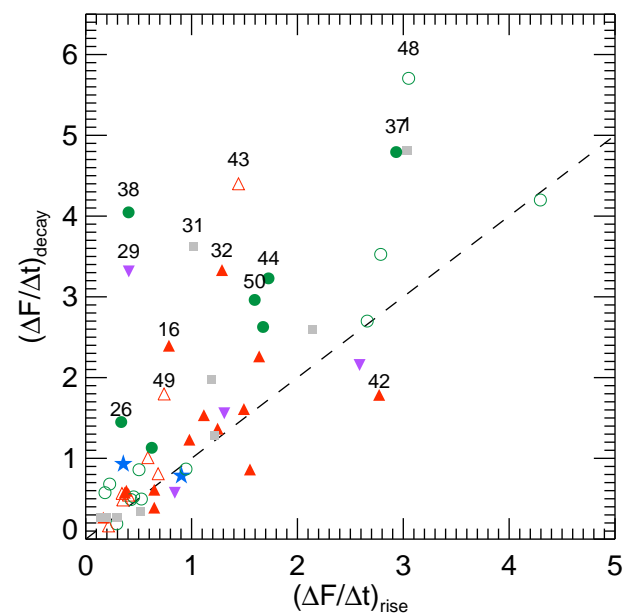}  \\       
  \caption{Fading versus rising flux rates, with the dashed line marking where the rates are equal. Flares numbers which deviate from this relation by more than 1\,$\sigma$ are marked.}    
 \label{fig:vrise_vfade}
 \end{figure}

 We have also computed the flare rise and decay rates, defined as ($\Delta$F/$\Delta$t)$_{rise}$ and ($\Delta$F/$\Delta$t)$_{decay}$. We show these rates  in Fig~\ref{fig:vrise_vfade}. Most of the flares seem to present similar rise and decay flux rates, but some flares show faster rates during decay than during rise. Some of these flares (flares 29, 31, 37, 38, and 44) also displayed decays to lower flux values than those measured before the flare onset, as described before (see Fig.~\ref{fig:rise_fade}). There are also several positive-lagged flares displaying larger decay than rising decays.

\subsection{Cross-correlation analysis}\label{sec:cross}
For all the optical flares identified in this work (see Figs.~\ref{fig:LCcomplete} and \ref{fig:LCcomplete2} and  Table~\ref{tab:flares}), we have performed a cross-correlation analysis with the X-ray light curves of the system in order to identify possible time lags between the optical and X-ray emission. For this analysis, we have used the discrete cross-correlation function (DCF, \citealt{edelson1988}), a method which avoids interpolating in the temporal domain when the temporal binning is not regular (this is the case for some of the optical observations in this work). In the results from the DCF analysis, a positive lag means that the optical follows the X-rays, a negative lag means that the optical precedes the X-rays, and zero lag means that both the optical and the X-rays are simultaneous.

\begin{figure}[h]
\begin{center}
   \includegraphics[width=8cm]{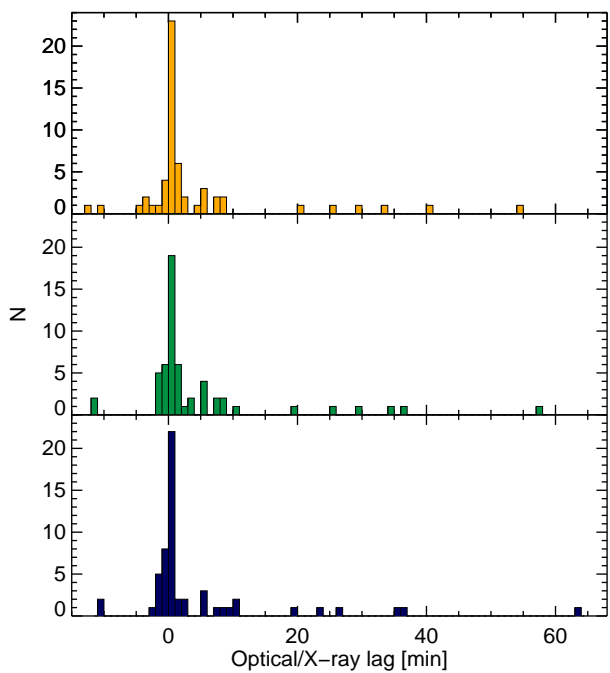}  \\       
 \vspace*{-0.3cm}

\end{center}

 \caption{Histogram of the measured lags between the optical and soft X-rays (3--10\,keV; top), optical and hard X-rays (20--80\,keV for \textit{INTEGRAL} data, 10--79\,keV for \textit{NuSTAR} data; middle), and optical and very hard X-rays (80--200\,keV; bottom). Bin size is 1\,min. }    
\label{fig:histo_lags}
\end{figure}

The results of DCF analysis for each flare are summarised in Table~\ref{tab:lags}. An histogram of the measured lags between the optical emission and the different X-ray bands is shown in Fig.~\ref{fig:histo_lags}. 

To estimate the uncertainties on the measured lags, we have performed 1000 Monte-Carlo simulations per flare, taking the optical and X-ray flux uncertainties and the temporal binning into account to randomly simulate each light curve. Considering the obtained uncertainties of the lags, in some parts of the outburst there could be optical lags up to some tens of seconds which we cannot measure with these data, while in other cases we are able to measure accurately lags of a few seconds.

In approximately a third of the flares we find that the optical emission is well correlated with the X-ray light curve and no lags (or lags smaller than the uncertainties) have been found. However, we have also found several flares displaying positive lags from tens of seconds to several minutes (the optical emission following the X-rays), and some negative lags of several minutes (with the optical emission preceding the X-rays).

\subsection{Hardness evolution}

\begin{figure}[h]
\begin{center}

  \includegraphics[width=7cm]{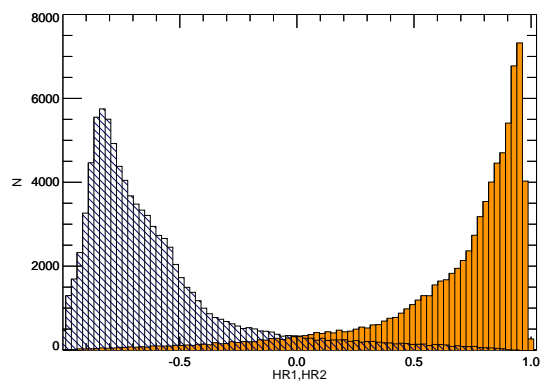}  \\       
\end{center}
 \vspace*{-0.3cm}

  \caption{Histogram of the hardness-ratios: HR1 (solid orange) and HR2 (dark blue striped). Bin size is 0.025. N is the number of observations within each hardness ratio bin.}  
\label{fig:histo_HRs}
\end{figure}

In order to complement our analysis and to search for spectral changes during the studied events, we have also analysed the evolution of the X-ray hardness-ratio (HR):
\begin{equation}
HR_{i} = \frac{CR_{a,i}-CR_{b,i}}{CR_{a,i}+CR_{b,i}}
\end{equation}

\noindent with $a,i$ and $b,i$ the two energy bands for which we have calculated each HR$_{i}$. We define HR1 as the ratio between the soft (3--10\,keV) and the hard (20--80\,keV for \textit{INTEGRAL}, 10--79\,keV for \textit{NuSTAR}) X-ray bands, and HR2 as the ratio between the hard (20--80\,keV) and very hard (80--200\,keV bands) for \textit{INTEGRAL}. A distribution of the values of these hardness-ratios is shown in Fig.~\ref{fig:histo_HRs}.

\subsection{Classification of the flares}

We have classified the flares according to the measured lags.

\subsubsection{Simultaneous optical and X-ray flares}\label{sub:simult}

\begin{figure*}[h]
   \includegraphics[width=9cm,height=0.22\textheight]{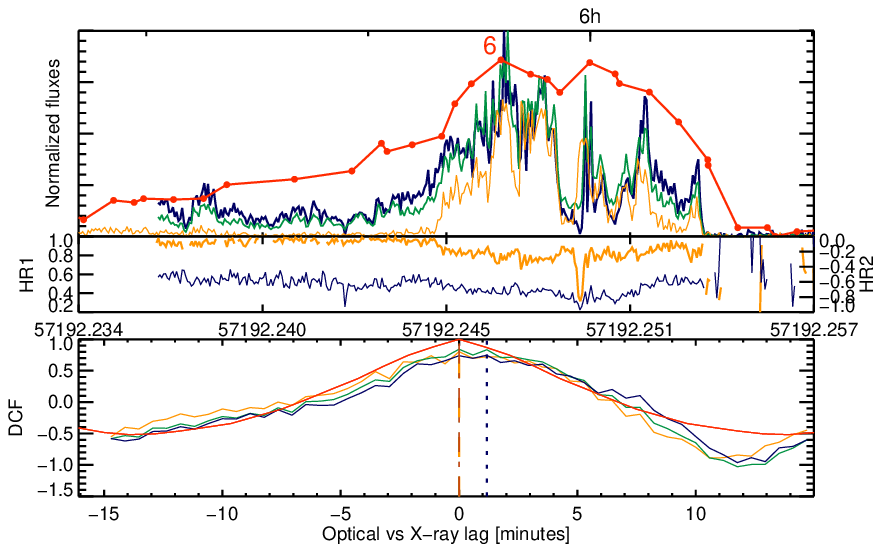}
   \includegraphics[width=9cm,height=0.22\textheight]{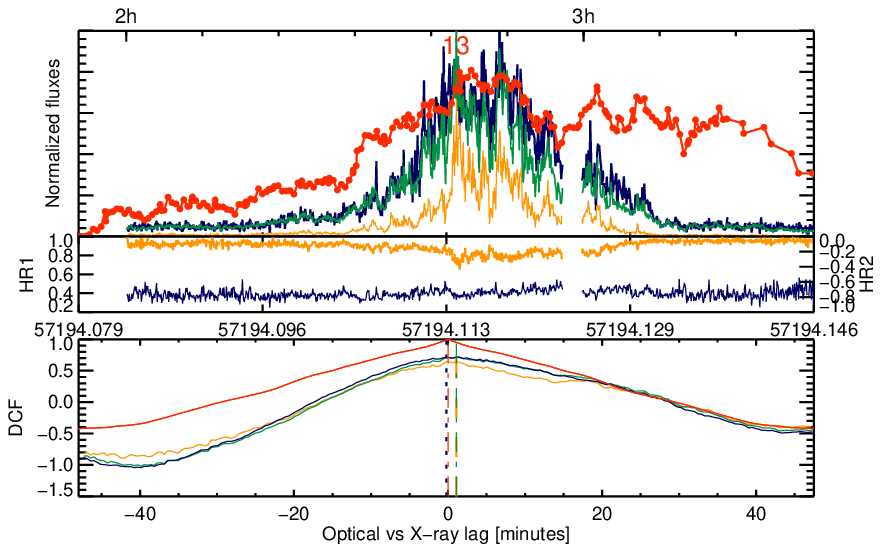}\\
\includegraphics[width=9cm,height=0.22\textheight]{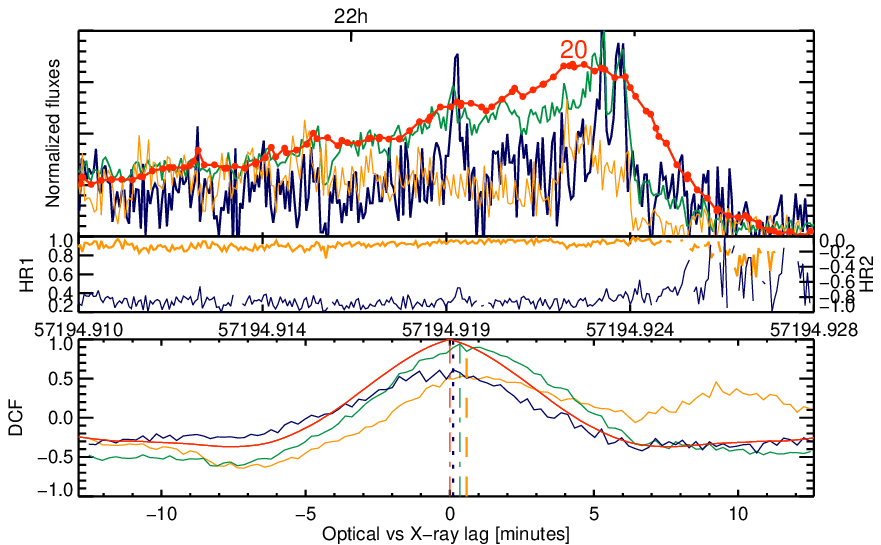}
\includegraphics[width=9cm,height=0.22\textheight]{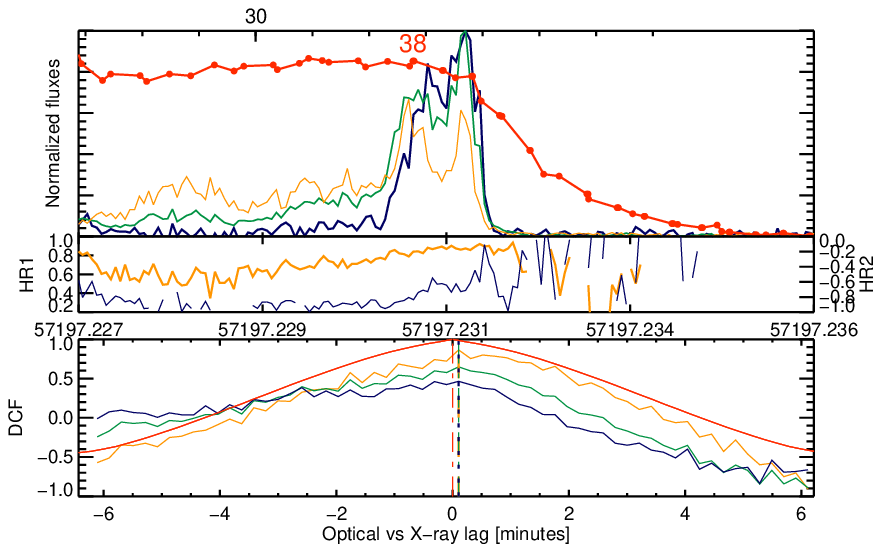}\\
\includegraphics[width=9cm,height=0.22\textheight]{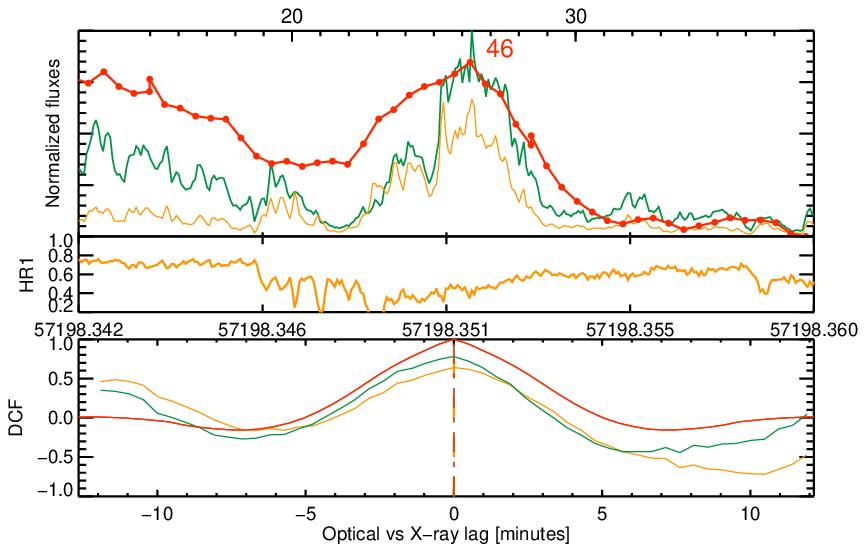}
\includegraphics[width=9cm,height=0.22\textheight]{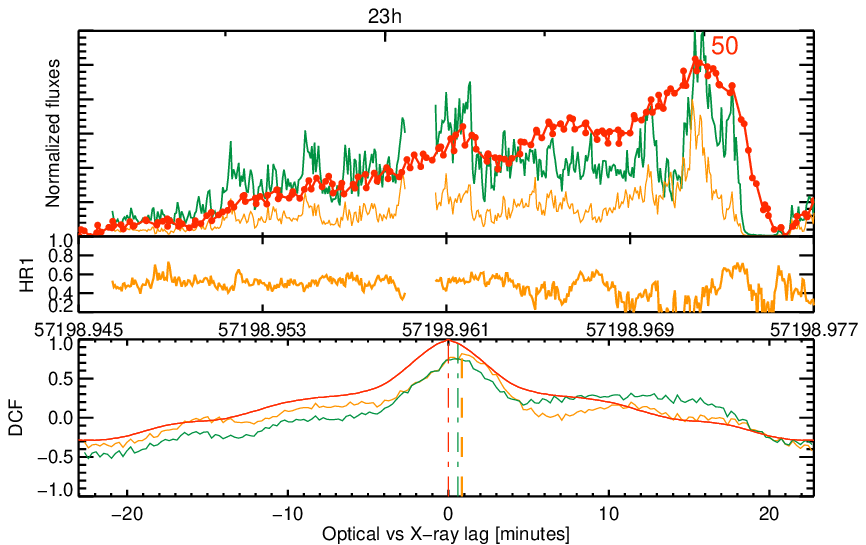}\\
\includegraphics[width=9cm,height=0.22\textheight]{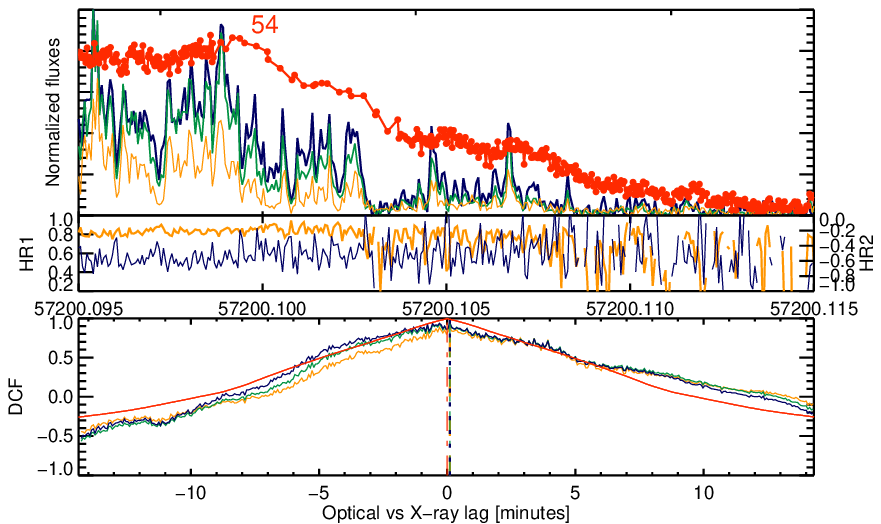}
\includegraphics[width=9cm,height=0.22\textheight]{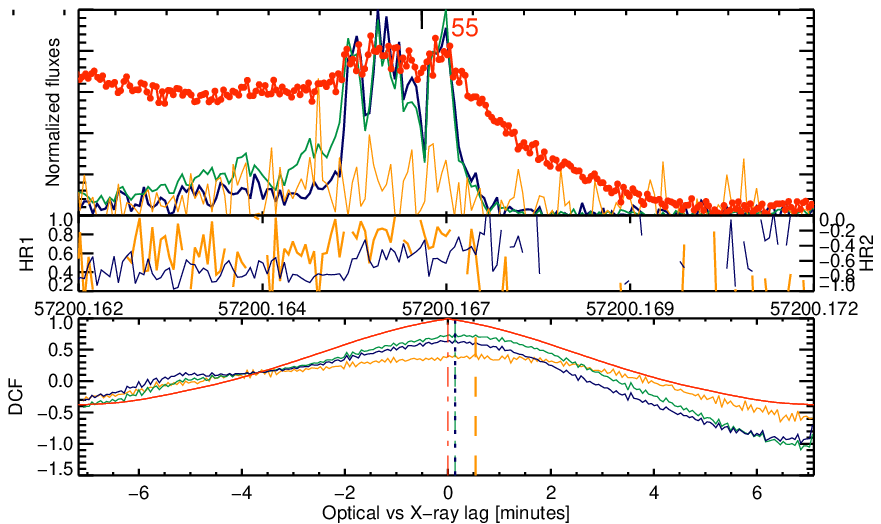}
\\
 \caption{Zoomed-in optical and X-ray light curves where optical flares are essentially simultaneous with X-ray flares (optical - red, soft X-ray (3-10keV) - orange, hard X-ray (20-80keV for ISGRI, or 10-79keV for \textit{NuSTAR}) - green, very hard X-ray (80-200keV) - dark blue). For each subplot, the HRs (HR1 - orange, HR2 - dark blue), and the DCFs (optical ACF - red, optical/soft X-ray DCF - orange, optical/hard X-ray DCF - green, optical/very hard X-ray - dark blue) are shown below the light curves. The variety of behaviour is discussed in the text.}    
\label{fig:simult}
\end{figure*}

For some flares, the optical and X-ray peaks do not show temporal lags within the uncertainties on the lag determination (see Table~\ref{tab:lags}). Some examples of simultaneous optical and X-ray flares are shown in Fig.\ref{fig:simult}. 

In these cases we observe two different flare patterns:
\begin{itemize}
 \item The optical flares seem to envelop the X-ray flares (see flares 6, 13, 46, and 54 in Fig.\ref{fig:simult}). 
 \item The optical and X-ray flare peaks are simultaneous, although the optical flares show longer decay times than the X-ray flares, which decay abruptly (see flares 20, 38, and 55 in Fig.\ref{fig:simult}). These flares are the asymmetric flares in Fig.~\ref{fig:rise_fade} mentioned in Sect.~\ref{sec:optflares}. This difference between the optical and X-ray flare profiles leads to an asymmetry in the DCF, which sometimes provides peak values slightly larger than zero. Other times we observe double flares displaying similar behaviour (flares 19--20) and also double symmetric structures in which each flare is composed of two sub-flares (flares 28--29 and 30--31). These regions are marked with light green boxes in Fig.~\ref{fig:LCcomplete}.
\end{itemize}

%\twocolumn

\subsubsection{Optical emission following the X-rays}\label{sub:X_opt}
In some flares we have observed that the optical emission is lagging the X-rays (positive lags). These lags are typically of several minutes, although time lags shorter than 2\,min have also been observed. We have chosen 2\,min as the limiting value to distinguish between short and long lags ($>$ 2 min), because as we will detail in the discussion, the size of the binary system is close to this value.

\begin{itemize}

\item \textbf{Short lags ($<$ 2 min)} 
For some flares we measure optical lags of several tens of seconds. Given the different monitoring frequency provided by the different optical telescopes, these results should be interpreted carefully. The uncertainties on the estimation of the lags takes this effect into account.

\begin{itemize}

\item \textit{Single flares:} some flares have been found to be delayed by 0.5--2\,min (e. g. flares 2, 43, 45, 49, and 53, see Fig.~\ref{fig:X_opt_short}). In this case, the optical flares display similar, although smoother shapes than the X-ray flares (contrary to the flares with slower optical fading described in Sect.~\ref{sub:simult}).

\begin{figure*}
   \includegraphics[width=9cm,height=0.24\textheight]{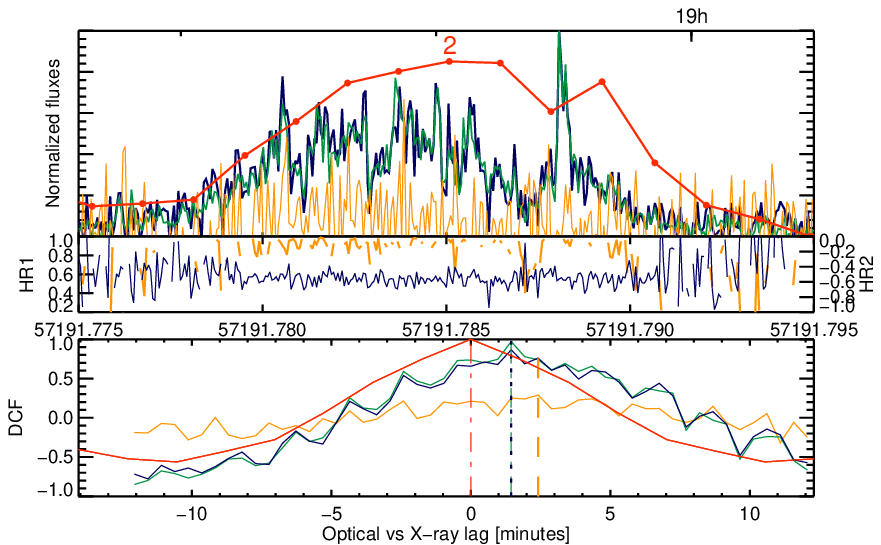}
    \includegraphics[width=9cm,height=0.24\textheight]{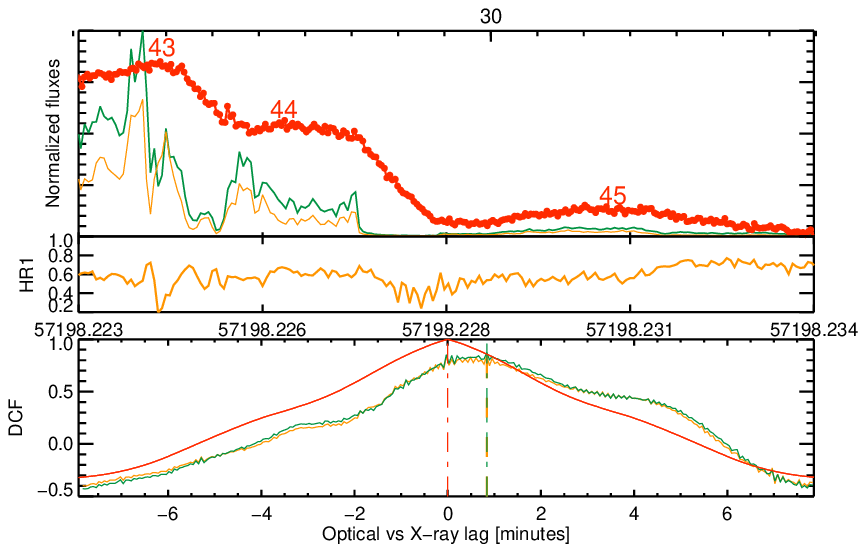}\\
     \includegraphics[width=9cm,height=0.24\textheight]{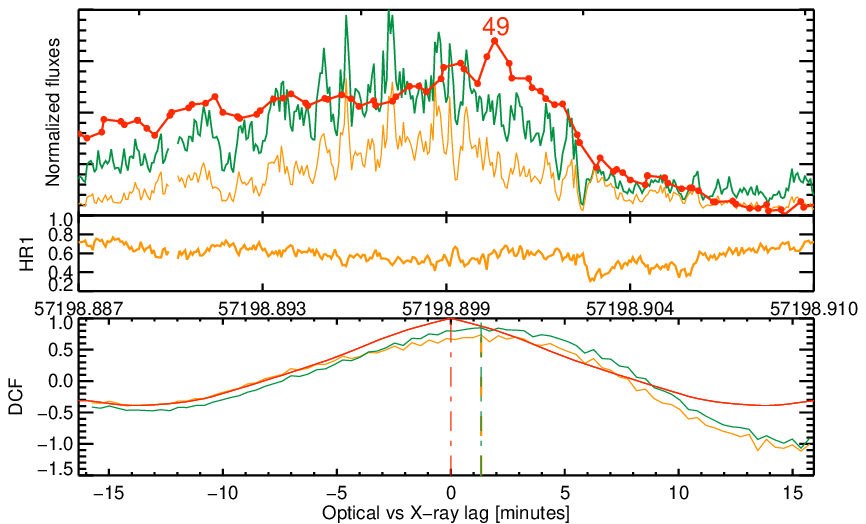} 
     \includegraphics[width=9cm,height=0.24\textheight]{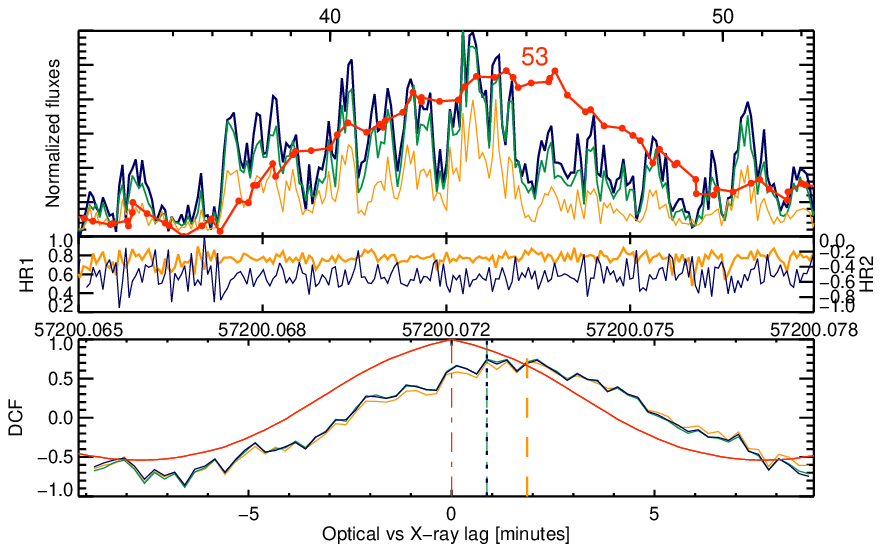}\\
      \caption{Same as in Fig.~\ref{fig:simult} but for flares with optical emission following the X-rays with short lags ($<$2\,min).}  
      \label{fig:X_opt_short}

    \includegraphics[width=9cm,height=0.24\textheight]{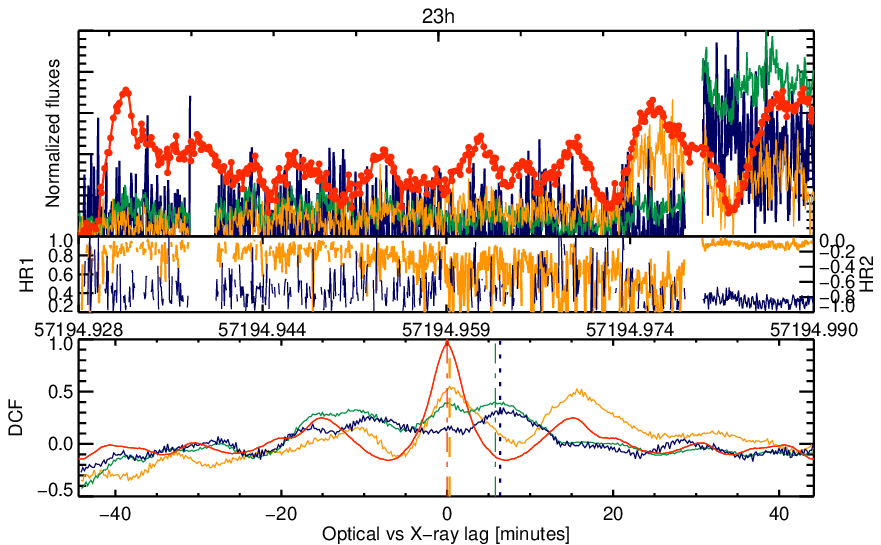}
\includegraphics[width=9cm,height=0.24\textheight]{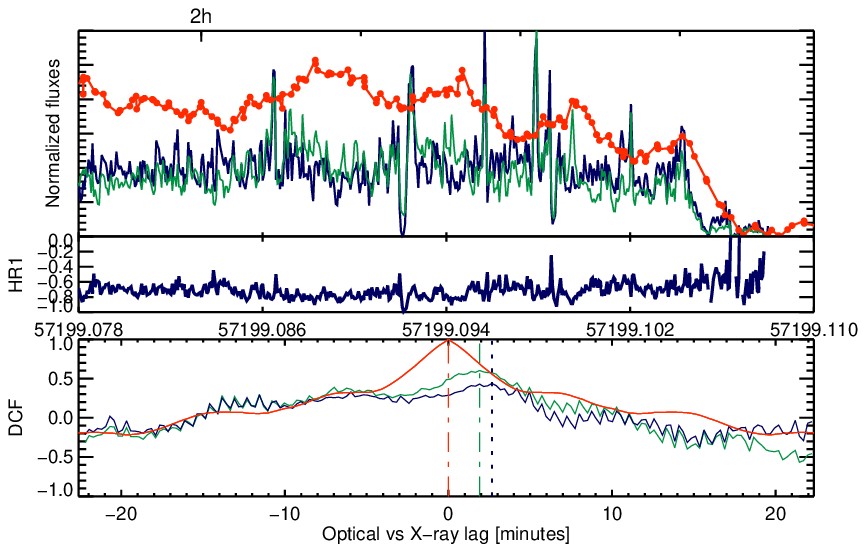}\\
\includegraphics[width=9cm,height=0.24\textheight]{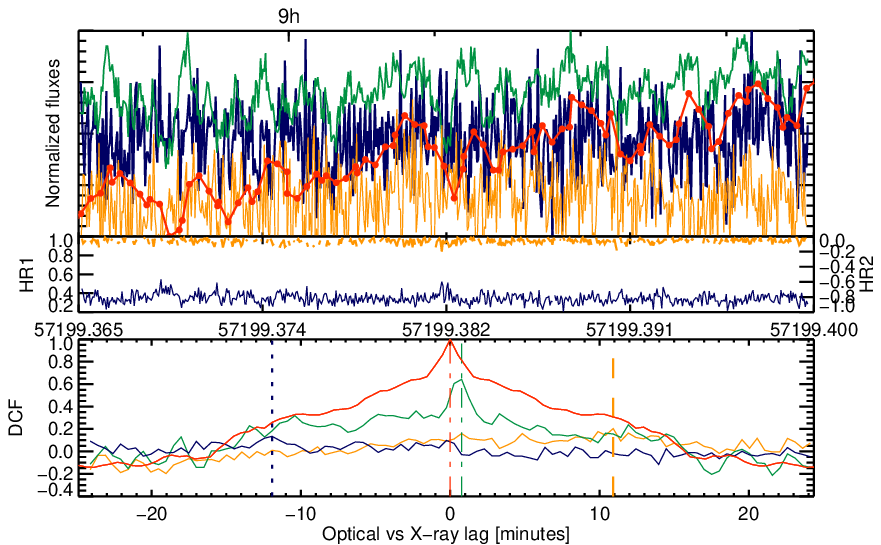}
   \includegraphics[width=9cm,height=0.24\textheight]{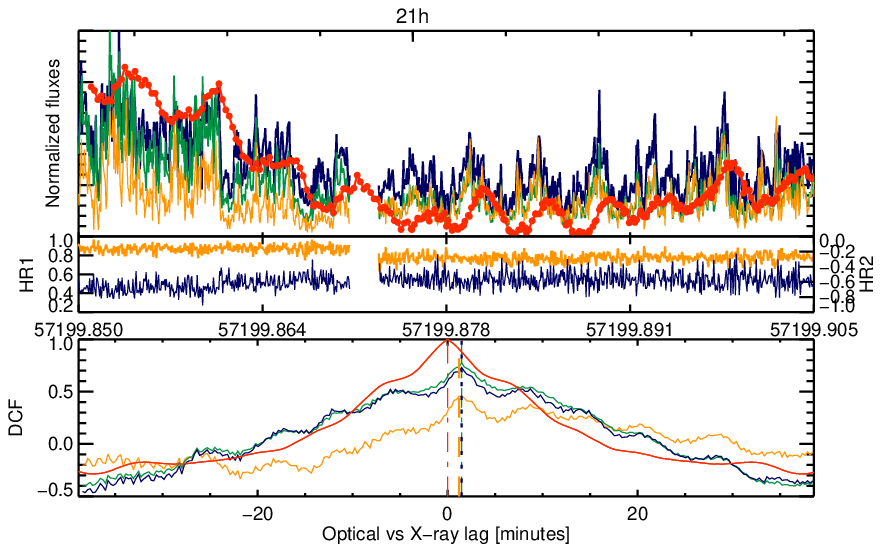}\\
 \caption{Same as in Fig.~\ref{fig:simult} but for some regions showing `heartbeat-type' oscillations.}    
\label{fig:flick}

\end{figure*}

\item \textit{'Heartbeat-type' oscillations:} in some epochs of the outburst, we have observed quasi-periodic oscillations (QPOs) in the optical and X-ray light curves (marked with purple boxes in Fig.~\ref{fig:LCcomplete2}). We show some examples of these `heartbeat-type' oscillations in Fig.~\ref{fig:flick}. In all cases the optical oscillations are obvious, but the X-ray oscillations, or at least the correlation with the optical and X-ray variations, are not so clear for all of them. In the last case (Fig.~\ref{fig:flick}, bottom right), the correlation between the optical and X-ray flickering is clear enough to perform a more detailed analysis. \linebreak 
\end{itemize}

\begin{figure*}
  \begin{center}
   \includegraphics[width=9cm,height=0.24\textheight]{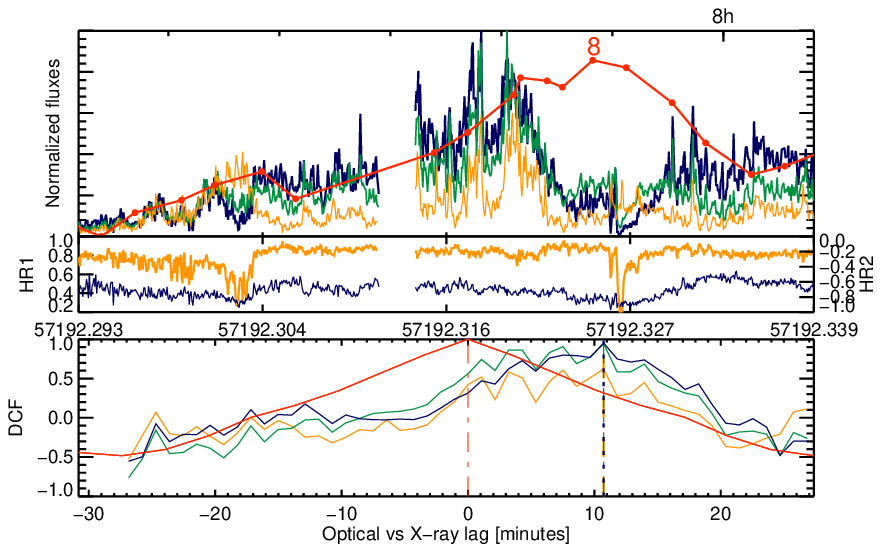}
   \includegraphics[width=0.5\textwidth,height=0.24\textheight]{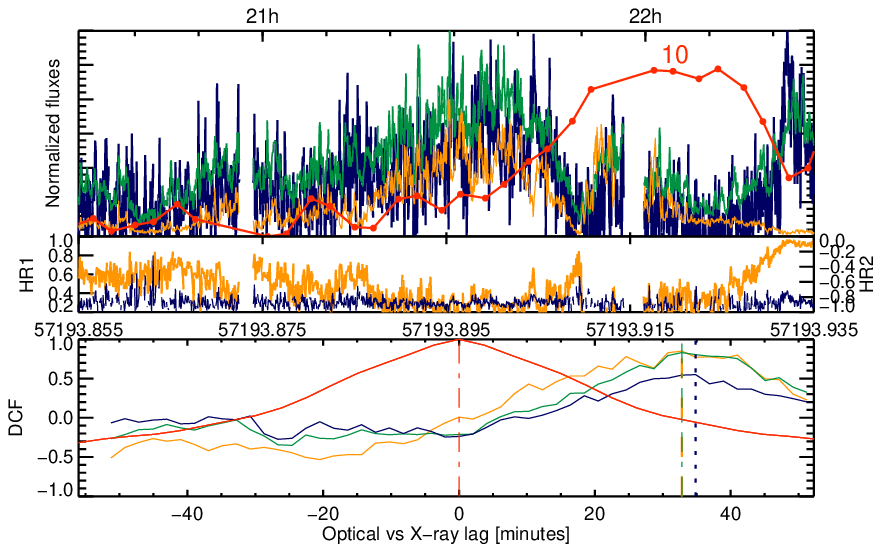}\\
\includegraphics[width=9cm,height=0.24\textheight]{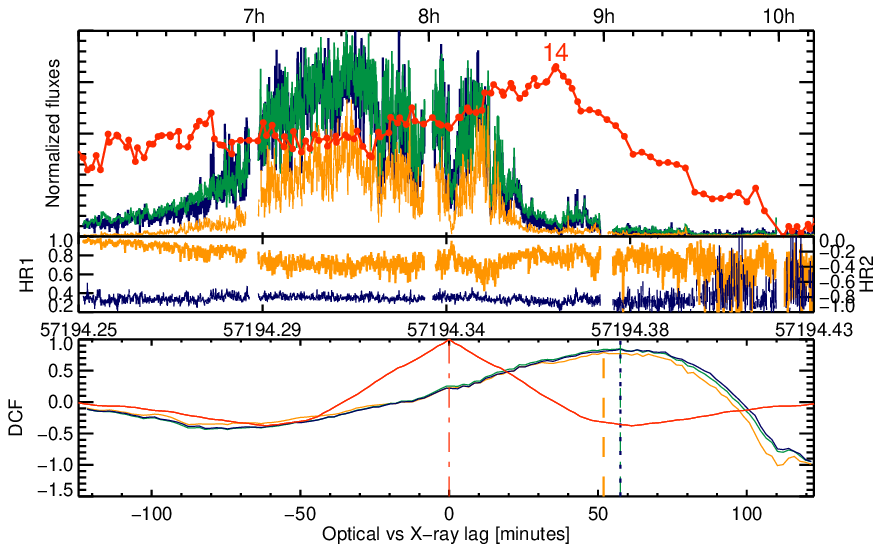}
\includegraphics[width=9cm,height=0.24\textheight]{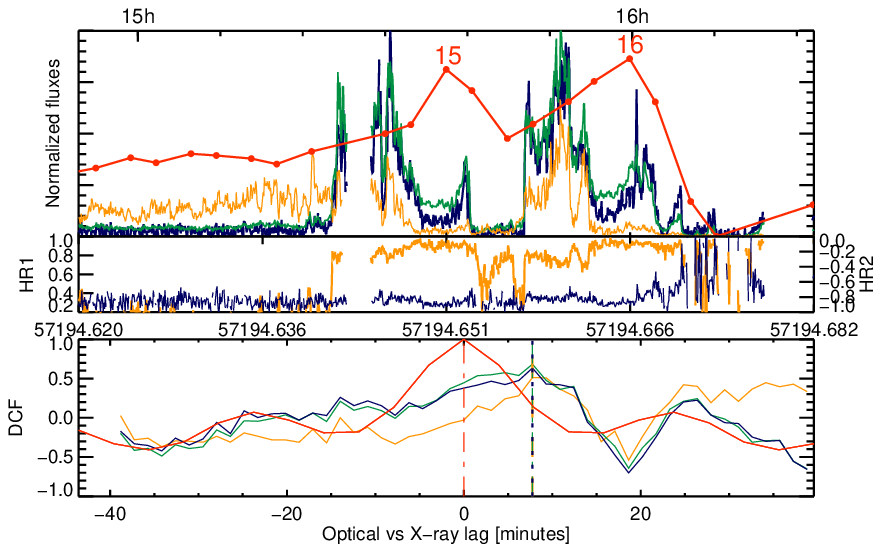}\\   
   \includegraphics[width=9cm,height=0.24\textheight]{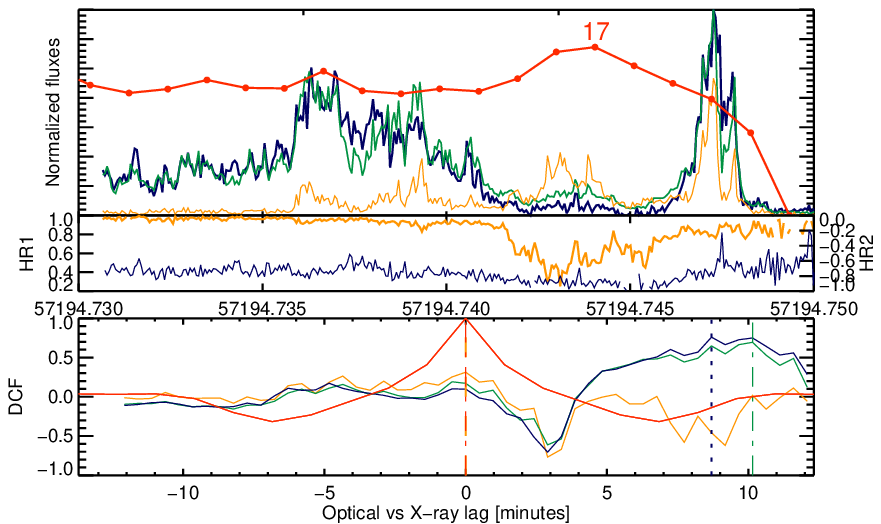}  
  \includegraphics[width=9cm,height=0.24\textheight]{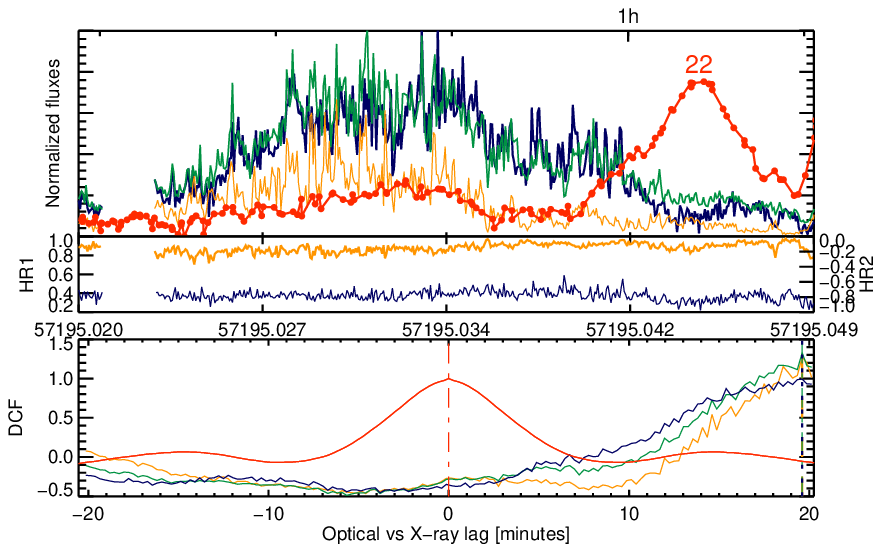}\\
   \includegraphics[width=9cm,height=0.24\textheight]{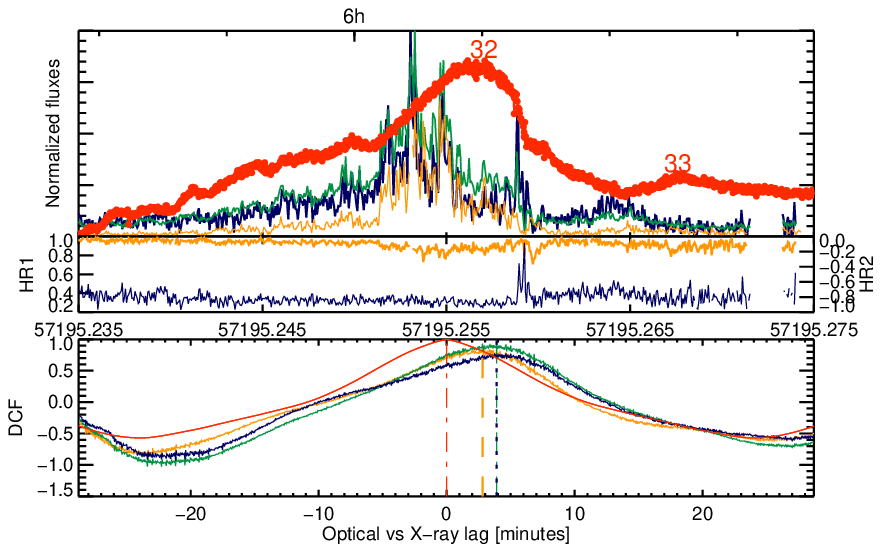}
   \includegraphics[width=9cm,height=0.24\textheight]{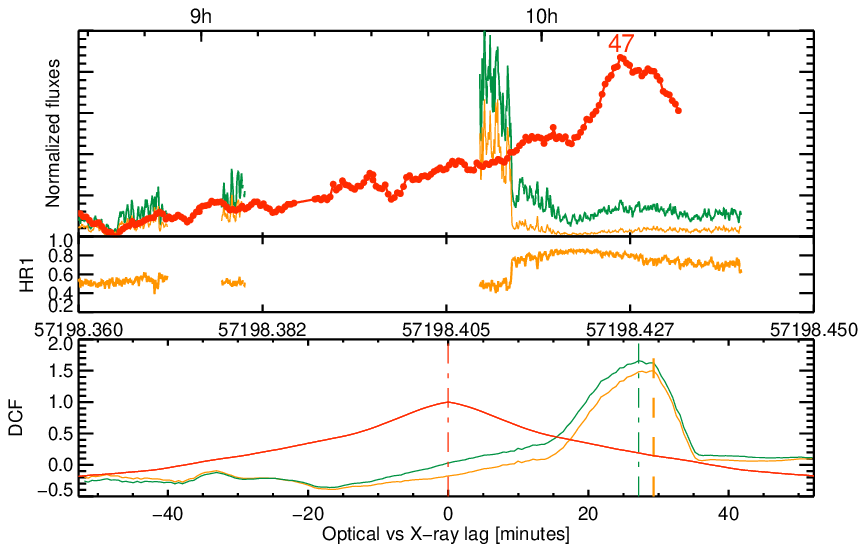}\\
      \end{center}
 \caption{Same as in Fig.~\ref{fig:simult} but for flares with optical emission following the X-rays with long lags ($>$2\,min).}    
\label{fig:X_opt}
\end{figure*}

\clearpage

\begin{table*}[htp]
\caption{Time delays measured from the DCF for each optical flare.}             % title of Table
\label{tab:lags}      % is used to refer this table in the text
 \begin{tabular}{r|r|cc|rrr|c}
\hline

Flare  & Time$_{peak}$  &T$_{ini}$ & T$_{fin}$ &Lag$_{opt/Xsoft}$ & Lag$_{opt/Xhard}$&Lag$_{opt/Xvery hard}$& Classification\\
\multicolumn{1}{c|}{\#} & \multicolumn{1}{r|}{(MJD-57190)} & \multicolumn{2}{c|}{(MJD-57190)}& \multicolumn{3}{c|}{[min]} & \\

\hline

       1 &   1.7657 $\pm$       0.0005 & 1.760 & 1.775 &       -0.1 $\pm$        0.9 &       -0.4 $\pm$        0.9 &       -0.4 $\pm$        0.8    & U    \\
       2 &   1.7851 $\pm$       0.0007 & 1.775 & 1.795 &        2.4 $\pm$        2.1 &        1.5 $\pm$        1.0 &          1 $\pm$          1    & P1       \\
       3 &   1.9722 $\pm$       0.0003 & 1.956 & 1.978 &        -2.4 $\pm$         1.4 &        0.6 $\pm$       0.5 &       0.0 $\pm$        0.5    & Z1       \\
       4 &   2.0616 $\pm$       0.0007 & 2.025 & 2.070 &        0.3 $\pm$        1.5 &       -0.6 $\pm$        0.9 &          -1$\pm$          1    & Z1       \\
       5 &   2.1737 $\pm$       0.0005 & 2.154 & 2.182 &        0 $\pm$         1 &        0.0 $\pm$             0.9 &      0.5 $\pm$        0.9    & Z1       \\
       6 &   2.2473 $\pm$       0.0004 & 2.234 & 2.257 &         0.2 $\pm$         1.0 &        0.2 $\pm$        1.0 &      0.2 $\pm$        1.0    & Z1       \\
       7 &   2.2824 $\pm$       0.0003 & 2.272 & 2.295 &        1.2 $\pm$        0.8 &         0.9 $\pm$         1.1 &         0$\pm$          1    & Z1       \\
       8 &    2.325 $\pm$        0.001 & 2.293 & 2.339 &        7.4 $\pm$        3.8 &        7.4 $\pm$        3.4 &       10.9 $\pm$        2.2    & P2       \\
       9 &    2.961 $\pm$        0.001 & 2.950 & 2.980 &                             &        1.6 $\pm$        2.5 &        0.3 $\pm$        2.2    & Z1       \\
      10 &    3.920 $\pm$        0.001 & 3.855 & 3.935 &       40.0 $\pm$        4.3 &       36.9 $\pm$        3.2 &       35.4 $\pm$        2.9    & P2       \\
      11 &    3.937 $\pm$        0.001 & 3.930 & 3.955 &       -1.8 $\pm$        0.8 &      -1.7 $\pm$         0.8 &       -1.7 $\pm$        0.8    & U       \\
      12 &   4.0034 $\pm$       0.0003 & 3.955 & 4.020 &      -10.3 $\pm$        0.4 &        7.8 $\pm$        0.4 &         24 $\pm$          2    & F       \\
      13 &   4.1150 $\pm$       0.0001 & 4.079 & 4.146 &        1.0 $\pm$        0.6 &        1.0 $\pm$        0.3 &        1.0 $\pm$        0.3    & P1       \\
      14 &   4.3629 $\pm$       0.0008 & 4.250 & 4.425 &       54.9 $\pm$        3.8 &       57.3 $\pm$        2.1 &       63.4 $\pm$        1.9    & P2       \\
      15 &    4.651 $\pm$        0.001 & 4.620 & 4.682 &        8.5 $\pm$        2.4 &        8.5 $\pm$        2.4 &        8.5 $\pm$        2.4    & P2       \\
      16 &    4.666 $\pm$        0.001 & 4.658 & 4.675 &       7.8 $\pm$        2.0 &        7.8 $\pm$         2.0 &        7.8 $\pm$        2.0    & P2       \\
      17 &   4.7441 $\pm$       0.0005 & 4.730 & 4.750 &        -4.4 $\pm$         1.0 &       10.2 $\pm$      0.9 &       10.2 $\pm$        0.8    & P2       \\
      18 &   4.8727 $\pm$       0.0002 & 4.858 & 4.885 &       -3.1 $\pm$        0.7 &      -11.1 $\pm$        0.4 &      -10.1 $\pm$        0.6    & N       \\
      19 &  4.90022 $\pm$      0.00008 & 4.886 & 4.906 &        1.7 $\pm$        0.2 &        1.5 $\pm$        0.2 &        0.3 $\pm$        0.2    & P1       \\
      20 &  4.92213 $\pm$      0.00008 & 4.910 & 4.928 &        0.4 $\pm$        0.2 &        0.4 $\pm$        0.1 &       -1.0 $\pm$        0.2    & Z2       \\
      21 &  5.00348 $\pm$      0.00005 & 4.990 & 5.015 &       0.20 $\pm$       0.09 &        3.0 $\pm$        0.2 &      -2.34 $\pm$       0.10    & F       \\
      22 &  5.04447 $\pm$      0.00007 & 5.010 & 5.049 &      19.23 $\pm$       0.08 &      19.23 $\pm$       0.08 &      19.23 $\pm$       0.05    &  P2      \\
      23 &  5.04946 $\pm$      0.00007 & 5.047 & 5.052 &        0.1 $\pm$        0.1 &       -0.3 $\pm$        0.1 &       -0.4 $\pm$        0.1    &  Z1      \\
      24 &  5.09850 $\pm$      0.00002 & 5.088 & 5.104 &       7.1 $\pm$        0.1 &         7.1 $\pm$        0.1 &        7.1 $\pm$        0.1    &  P2    \\
      25 &  5.10693 $\pm$      0.00002 & 5.101 & 5.116 &        0.0 $\pm$        0.1 &        0.0 $\pm$        0.1 &      -0.12 $\pm$       0.09    &  U      \\
      26 &  5.11048 $\pm$      0.00002 & 5.108 & 5.114 &       0.04 $\pm$       0.04 &       -0.1 $\pm$        0.2 &       -0.1 $\pm$       0.07    &  Z2      \\
      27 &  5.16931 $\pm$      0.00002 & 5.160 & 5.180 &       -0.1 $\pm$        0.2 &      -11.1 $\pm$        0.3 &      -10.8 $\pm$        0.5    &  N      \\
      28 &  5.21332 $\pm$      0.00002 & 5.205 & 5.220 &        0.1 $\pm$        0.2 &        0.0 $\pm$        0.2 &       -0.7 $\pm$        0.3    &  F      \\
      29 &  5.21603 $\pm$      0.00002 & 5.205 & 5.220 &        0.4 $\pm$        0.2 &        0.0 $\pm$        0.2 &       -0.3 $\pm$        0.3    &  F      \\
      30 &  5.22508 $\pm$      0.00002 & 5.218 & 5.227 &        0.3 $\pm$        0.6 &        0.5 $\pm$        0.3 &        -0.7 $\pm$       0.3    &  Z1      \\
      31 &  5.22837 $\pm$      0.00002 & 5.226 & 5.233 &       -1.0 $\pm$        0.6 &       -1.1 $\pm$        0.3 &       -1.1 $\pm$        0.3    &  U      \\
      32 &  5.25704 $\pm$      0.00002 & 5.231 & 5.264 &        2.8 $\pm$        0.9 &        3.4 $\pm$        0.4 &        4.1 $\pm$        0.5    &  P2      \\
      33 &  5.26760 $\pm$      0.00002 & 5.263 & 5.274 &        4.6 $\pm$        0.1 &        5.2 $\pm$        0.4 &        5.9 $\pm$        0.1    &  P2      \\
      34 &   7.0287 $\pm$       0.0002 & 7.007 & 7.055 &        0.0 $\pm$        0.3 &        1.3 $\pm$        0.5 &        5.7 $\pm$        0.3    &  U      \\
      35 &   7.2019 $\pm$       0.0002 & 7.155 & 7.215 &       33.9 $\pm$        0.8 &         34 $\pm$           1&       36.8 $\pm$        0.6    &  P2      \\
      36 &  7.22212 $\pm$      0.00010 & 7.190 & 7.231 &       25.1 $\pm$        0.5 &       25.7 $\pm$        0.5 &       26.0 $\pm$        0.7    &  P2      \\
      37 &  7.22549 $\pm$      0.00010 & 7.223 & 7.228 &       -0.1 $\pm$        0.4 &       -0.1 $\pm$        0.1 &        0.1 $\pm$        0.4    &  Z2      \\
      38 &  7.23110 $\pm$      0.00008 & 7.227 & 7.236 &        0.2 $\pm$        0.1 &        0.2 $\pm$        0.2 &        0.4 $\pm$        0.3    &  Z2      \\
      39 &   7.3121 $\pm$       0.0001 & 7.309 & 7.319 &        1.1 $\pm$        0.2 &        2.2 $\pm$        0.5 &        2.2 $\pm$        0.4    &  U      \\
      40 &    7.368 $\pm$        0.001 & 7.338 & 7.375 &      -12.1 $\pm$        0.7 &       -1.3 $\pm$        0.7 &       -1.5 $\pm$        0.7    &  U      \\
      41 &  8.21768 $\pm$      0.00002 & 8.214 & 8.220 &        0.0 $\pm$        0.3 &        0.0 $\pm$        0.2 &                                &  Z1      \\
      42 &  8.22090 $\pm$      0.00002 & 8.214 & 8.223 &        5.0 $\pm$        0.3 &        5.3 $\pm$        0.2 &                                &  P2      \\
      43 &  8.22425 $\pm$      0.00002 & 8.223 & 8.228 &        0.8 $\pm$        0.2 &        1.0 $\pm$        0.3 &                                &  P1      \\
      44 &  8.22607 $\pm$      0.00002 & 8.223 & 8.228 &        0.4 $\pm$        0.2 &        0.4 $\pm$        0.3 &                                &  Z2      \\
      45 &  8.23099 $\pm$      0.00002 & 8.228 & 8.234 &        0.5 $\pm$        0.1 &        0.4 $\pm$        0.1 &                                &  P1      \\
      46 &   8.3516 $\pm$       0.0002 & 8.342 & 8.360 &        0.1 $\pm$        0.3 &       -0.1 $\pm$        0.4 &                                &  Z1      \\
      47 &   8.4264 $\pm$       0.0002 & 8.360 & 8.450 &        29.1 $\pm$       1.3 &        29.1 $\pm$       1.2 &                                &  P2      \\
      48 &   8.8830 $\pm$       0.0002 & 8.880 & 8.886 &        0.2 $\pm$        0.2 &        0.2 $\pm$        0.2 &                                &  Z1      \\
      49 &   8.9000 $\pm$       0.0001 & 8.887 & 8.910 &        1.3 $\pm$        0.2 &        1.3 $\pm$        0.3 &                                &  P1      \\
      50 &  8.97187 $\pm$      0.00008 & 8.945 & 8.977 &        0.8 $\pm$        0.2 &        0.8 $\pm$        0.2 &                                &  Z2      \\
      51 &  9.98513 $\pm$      0.00005 & 9.968 & 9.995 &                             &        0.7 $\pm$        0.1 &        0.7 $\pm$        0.1    &  P1      \\
      52 &  10.03205 $\pm$      0.00005 & 10.021 & 10.043 &       2.0 $\pm$         0.1 &        1.5 $\pm$        0.1 &        1.5 $\pm$        0.3    &  P1      \\
      53 &  10.07307 $\pm$      0.00006 & 10.065 & 10.078 &       0.9 $\pm$         0.1 &        0.8 $\pm$        0.1 &        0.8 $\pm$        0.1    &  P1      \\
      54 &  10.09927 $\pm$      0.00003 & 10.095 & 10.115 &        0.1 $\pm$        0.1 &        0.10 $\pm$        0.07 &        0.10 $\pm$       0.07    &  Z1      \\
      55 &  10.16686 $\pm$      0.00002 & 10.162 & 10.172 &        0.5 $\pm$        0.5 &        0.1 $\pm$        0.2 &        0.1 $\pm$        0.2    &  Z2      \\
      56 &   10.2064 $\pm$       0.0008 & 10.195 & 10.218 &       0.0 $\pm$        2.3 &       -1.0 $\pm$         1.5 &       -1.0 $\pm$        1.5    &  Z1      \\

\\
\hline 

\end{tabular}

\begin{flushleft} Flare classification according to the measured lags: Zero-lags in smooth flares (Z1), zero-lags with higher flux decay than rise rates and lower end fluxes (Z2), positive short lags (P1), positive long lags (P2), negative lags (N), frequency dependent lags (F), uncertain (U).\\
%inserts single line
\end{flushleft}
\end{table*}
\clearpage

\noindent Between MJD~57199.850 and 57199.875 the average optical and X-ray fluxes were decaying, while from MJD~57199.875 to MJD~57199.905 the average optical and X-ray fluxes remained roughly constant. We have calculated the period of the QPO in both epochs using Lomb-Scargle techniques \citep{lomb1976,scargle1982}, obtaining values of 8.1\,min for the decay phase and 6.4\,min for the stable phase, with very similar results in the optical and X-ray light curves. We derive optical lags of 1.2$\pm$0.3\,min, 1.5$\pm$0.3\,min, and 1.5$\pm$0.3\,min when comparing with the soft, hard, and very hard energy bands, respectively.

No significant changes have been observed in HR1 and HR2 in the single and `heartbeat-type' optical flares displaying short delays.

\item \textbf{Long lags ($>$ 2\,min)}. The epochs where positive lags larger than 2\,min have been measured are marked with brown continuous boxes in Fig.~\ref{fig:LCcomplete} and \ref{fig:LCcomplete2}. Closer examination of some representative flares showing these positive lags is displayed in Fig.~\ref{fig:X_opt}. 
\begin{itemize}

\item \textit{Single-peaked flares:} examination of the optical and X-ray light curves shows that, for most of the time, we can identify one single optical flare lagging the X-ray flare (see Fig.~\ref{fig:X_opt}). The delayed optical flares tend to show symmetric and quasi linear brightening and fading rise and decays (flares 15--16, 17, 22, and 47) while in a few cases, the optical flares display smoother peaks (flares 10,32).

In many cases the optical flare begins when the X-ray flare has already decayed (e.g. flares 15--16, 17, 22, 47). On other occasions, the optical flare begins during the X-ray flare (flares 8, 10). In most cases the X-ray and optical flares occur after a longer-term continuous increase in the optical (e.g. flares 10, 14 15, 16, 17, 32, and 47).

\item \textit{Multi-flare complex regions:} Sometimes, multiple optical and X-ray flares are observed, and identifying which optical flare is related with an X-ray flare is not easy. In these cases, we cannot select an interval containing only one optical and one X-ray flare, and the DCF results can be affected by the presence of other flares.

\begin{figure}[h]
\begin{center}

   \includegraphics[width=9cm]{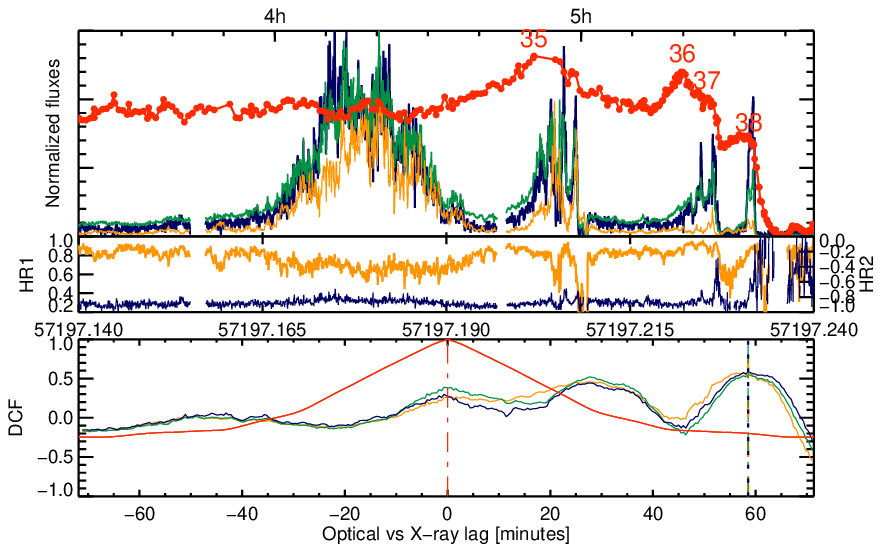}\\
      \includegraphics[width=9cm]{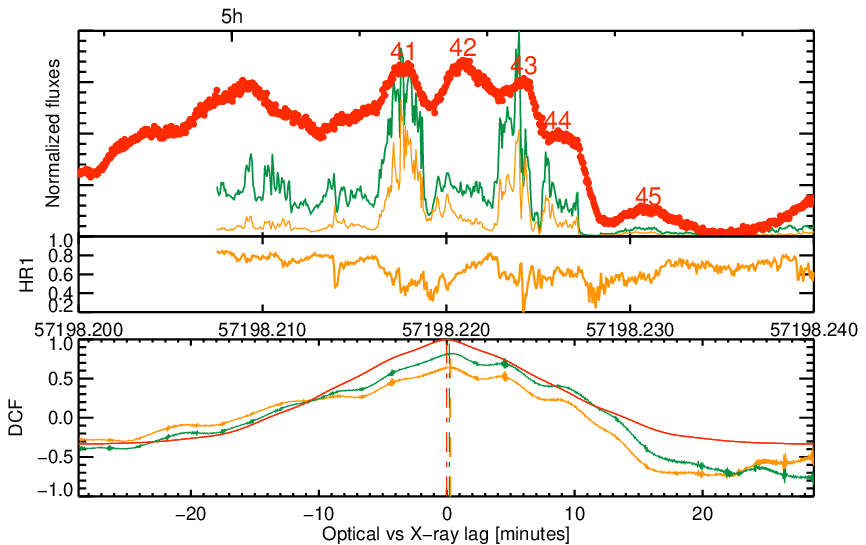}    

 \end{center}

 \caption{Same as in Fig.~\ref{fig:simult} but for complex cases in which there are several optical flares which follow the X-rays.}    
\label{fig:X_opt_complex}
\end{figure}

One of these regions is that from MJD~57197.140--57197.240 (see Fig.~\ref{fig:X_opt_complex}, top), where the DCF of this region gives two peaks at $\sim$58.5 and $\sim$27~min and it is hard to be sure which optical flare is related with each X-ray flare. 
If we consider that flares 35--36 are related to the first two X-ray flares, and calculate the DCF for shorter regions trying to avoid the presence of the other flares, we obtain time-delays of $\sim$34\,min for flare 35, and $\sim$25~min for flare 36 (see Table~\ref{tab:lags}). The sum of both time delays would be in agreement with the first DCF peak at $\sim$59\,min. After flares 35--36, two optical flares (37--38) occur with simultaneous X-ray emission. Both display very fast X-ray and slower optical decays.

Another peculiar region is that from MJD~57198.214 to 57198.229, where several optical and X-ray flares appear (see Fig.~\ref{fig:X_opt_complex}, bottom). In this case, the DCF gives the main result at close to zero, but another two peaks at $\sim$5 and 9~min are significant. Indeed, if we calculate the DCF in a region including only flares 41 and 42, the DCF peaks at $\sim$5\,min, which is the distance between both optical flares.
While flares 41 and 43 can be related with two simultaneous X-ray flares, flares 42 and 44 seem to be related with the two fainter X-ray flares. On the other hand, flares 41 and 42 are symmetric and display similar shape in the optical, while flares 43 and 44 display large optical flux decays with longer optical than X-ray decay times (similarly to flares 37--38).

\end{itemize}

The HR evolution of these long-delayed optical flares displays different patterns. Sometimes we observe a sudden hardening in HR1 just before the optical flares, at the beginning (see flares 8, 15--16) or just after the end of the previous X-ray flares (flares 36, and 47). In other cases, the optical brightening takes place during a smooth softening of the X-ray emission (as for flares 10, 14, and 17). In many cases a fast X-ray flare takes place after a delayed optical flare. These flares are harder than most of the flares observed in the outburst and can be identified by spikes in the HR2 evolution occurring several minutes after the optical flares (we have observed these spikes after flare 17, 22, 29, 32, 35, 36, 40) . Some of these fast flares show also simultaneous optical emission and have been identified as individual flares in this work (flares 23, 37, 38).

\end{itemize}

\begin{figure}[h]
  \begin{center}
    \includegraphics[width=9cm]{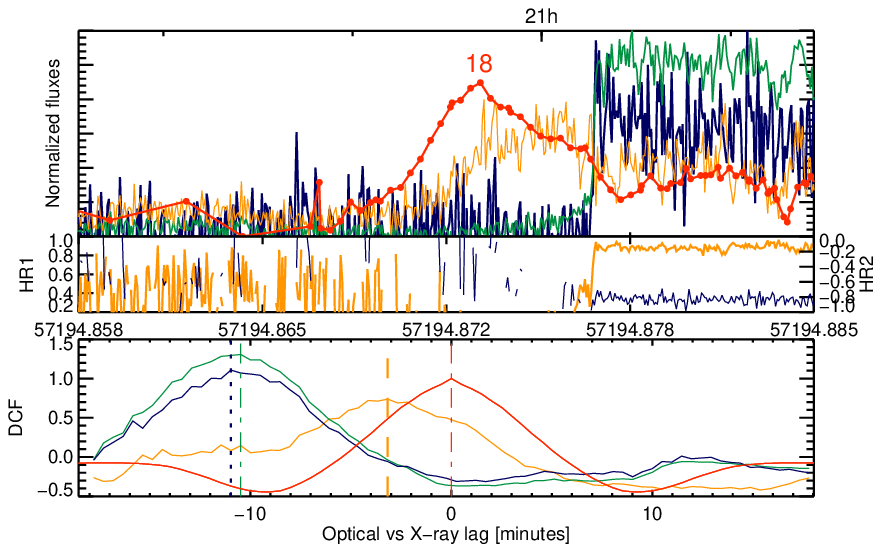}\\
    \includegraphics[width=9cm]{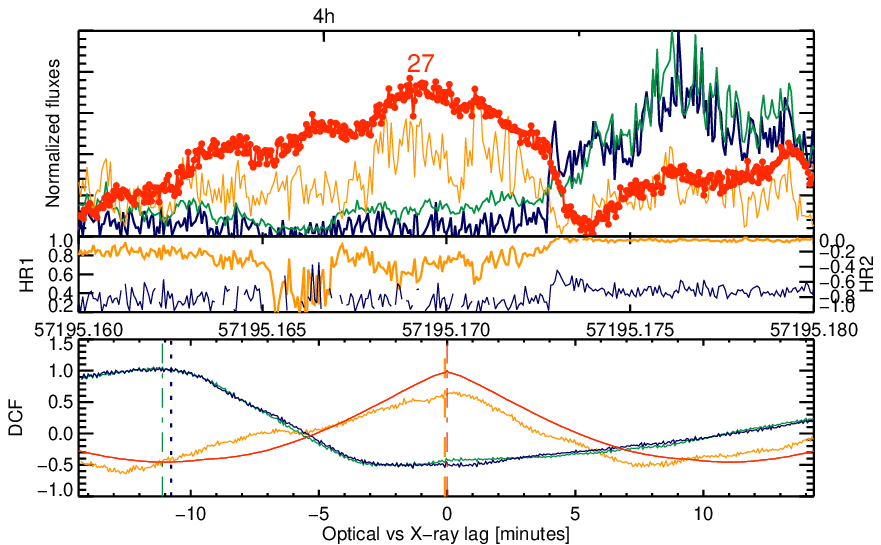}\\
      \end{center}
      \caption{Same as in previous figures, but for cases in which the optical and soft X-ray emission precedes the hard X-rays.}
      \label{fig:opt_X}
\end{figure}

\subsubsection{Optical emission leading the X-rays}\label{sub:opt_X}
We have identified several regions in which the optical emission and/or soft X-ray activity precedes the brightening in hard X-rays. These intervals are marked with blue boxes in Fig.~\ref{fig:LCcomplete} and \ref{fig:LCcomplete2}. We show two examples (flares 18, 27) in Fig.~\ref{fig:opt_X}. In these flares we measure negative lags of $\sim$-11\,min, between the optical and hard X-ray light curves and lags of $\sim$-3.1 and -0.1\,min between the optical and the soft X-rays. There are two epochs around MJD~57194.630 and MJD~57195.430 (also marked with blue boxes in  Figs.~\ref{fig:LCcomplete} and \ref{fig:LCcomplete2}), where a brightening in soft X-rays precedes by several minutes a sudden increase of the hard X-ray light curve. Unfortunately there are no optical data during these two epochs.

In all these flares we can identify two clear epochs in the optical and X-ray light curves, with the transition between them characterized by a sudden brightening of the hard X-ray light curve and a sudden hardening of the system light curves, as measured by the hardness ratio HR1 (see Fig.~\ref{fig:opt_X}).

\subsubsection{Frequency dependent lags}
\begin{figure}[h]

\includegraphics[width=9cm]{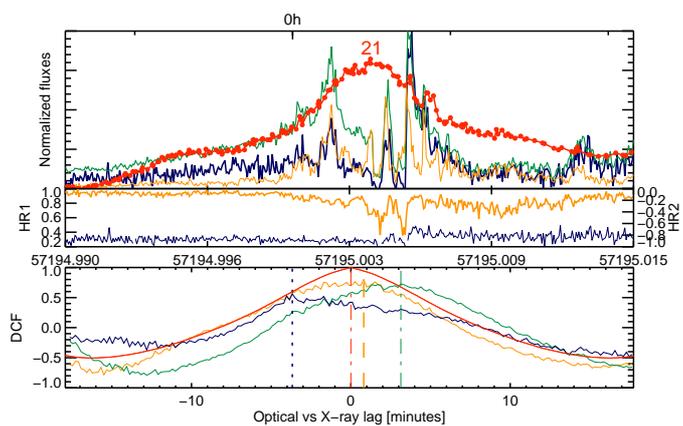}%
\caption{Same as in Fig.~\ref{fig:simult} but for a flare with energy-dependent optical/X-ray lags.}    
\label{fig:X_opt_complex2}
\end{figure}

We have also observed that on occasion the measured lag is positive or negative depending on the X-ray energy band we compare with. This is the case for flare 21, which displays a smooth peak in the optical, but during which several (at least 3) X-ray flares are detected (see Fig.~\ref{fig:X_opt_complex2}). As these three flares display different peak fluxes in the different X-ray bands, different lags are obtained depending on the energy band which we cross-correlate with, as the DCF identifies  different times for the peaks of each X-ray flare. Positive lags of $\sim$0.2 min and 3.01\,min are obtained when we correlate the optical with the soft and hard X-ray light curves, but a negative lag of -2.34\,min is obtained for the optical to very hard X-ray correlations. This is probably an effect due to changes in the hardness-ratio. Similar behaviour is observed in flares 28, 29.

\section{Discussion}
We have studied the optical and X-ray light curves of V404~Cyg during the period 18--28 June 2015, when intense flaring activity was observed. We have identified those flares observed in both energy ranges. We have used the DCF technique to search for time lags between the observed optical and X-ray variability.

\subsection{Flares displaying no time lags}
For a large fraction of the flares analysed in this work we do not find time lags between the optical and X-ray light curves (taking the uncertainties in the determination of the lags into account; see Fig. \ref{fig:histo_lags}). This suggests that, in these flares, the optical and X-ray emission are produced in the same, or at least in physically very close regions in the system, which can include emission from the inner hot flow (the corona or the base of the jet).

For V404~Cyg, the contribution of the jet synchrotron emission to the optical is controversial. It has been found to reach extreme values compared to other BHXBs which exhibit an optically thick to optically thin jet synchrotron break at lower frequencies \citep{russell2013}. \citet{maitra2017} found evidence that the break frequency was higher than the optical frequencies during their observations at $\sim$MJD~57200. However, \citet{russell2013} measured $\nu_{b} \sim$ 1.8 $\times 10^{14}$\,Hz during the 1989 outburst. \citet{koljonen2015} studied the relation between the photon index of the X-ray continuum and the break frequency for BHXBs and AGNs, assuming that the X-ray emission comes from the coronal power-law. From their analysis, it can be inferred that, for the hardest power law indexes observed during the 2015 outburst, a break frequency of $\nu_{b}\sim$10$^{15}$--10$^{16}$\,Hz can be reached. This implies that the break frequency can vary during the outburst, and the optical contribution from the jet could be significant for the hardest events. Fast variations in the break frequency have been observed in the past for GX~339--4 by \citet{gandhi2011}, who observed changes on timescales shorter than $\sim$\,1.5 hours.

We have identified a sub-class of no-lagged flares, which display a characteristic profile. In these flares, the optical and X-ray light curves peak simultaneously, but the decay of the X-ray light light curve is much faster than the optical light curve decay. Simultaneous infrared, optical, and X-ray observations during one of these flares (flare 44; see Fig.~\ref{fig:X_opt_complex}, bottom) were analysed by \citet{dallilar2017}. They measured the infrared, optical, and X-ray decay times of this flare and concluded that they were compatible with a cooling effect mainly due to a non-thermal mechanism.  This would likely be synchrotron emission, revealing that the corona or base of the jet was contributing to the optical emission at that epoch. We propose that the flares we have observed with similar profiles could also be produced by the same mechanism.

\subsection{Flares displaying positive time lags}
We have classified the positive-lagged flares in two categories: short- ($<$2\,min) and long- ($>$2\,min) lagged flares. Optical flares displaying short lags are compatible with reprocessing of incident X-rays (emitted close to the BH) in the accretion disc, and/or in the companion star. However, longer lags would require a different explanation. We have selected 2\,min as a limiting value, because it would correspond roughly to the light crossing time of the physical size of the binary system. From the known orbital parameters (see Sect.~\ref{sec:int}) we can infer the mean distance from the BH to the companion star, $a\sim$2.2$\times$10$^{12}$\,cm ($\sim$75\,light--s), and the maximum size of the disc, $r_{max}\sim$1.4$\times$10$^{12}$\,cm ($\sim$~45~light--s). When reprocessing of X-rays in the accretion disc and in the companion star is the mechanism producing the optical emission, we would observe positive lags due to light travel times and smoother optical than X-ray flare profiles, since the optical emission would be the sum of the contribution from multiple regions along the disc. To understand the observed lags, the inclination of the system needs to be considered. In a simplified model, the measured lags will  vary between zero lag to r$_{disc}$\,(1 + $\sin i$), where r$_{disc}$ is the disc radius in light seconds and $i$ the binary system inclination. The response from the companion star would oscillate within the range $a$ (1$\pm \sin i$) along the binary orbit, with the binary separation $a$ expressed in light seconds \citep{obrien2002, hynes2010}. The relative contribution of the companion response with respect to the disc would depend on the relation between the accretion disc and the companion star radii, but also on the inclination of the system and the thickness of the accretion disc which define what fraction of the companion star is covered. For V404~Cyg, with an inclination of $\sim$\,60 degrees\footnote{Quiescent studies of V404 Cyg have yielded estimates of the binary inclination in the range 56--67 degrees \citep{shahbaz1994, sanwal1996}, with systematic differences likely due to residual accretion disc contamination, as discussed by \citet{charles2018}.  Hence we use a value of $\sim$60 degrees here.}, lags up to $\sim$84\,s (1.4\,min) would be expected from reprocessing in the accretion disc, and lags between $\sim$10 and 140\,s (0.2--2.3\,min) would be expected when reprocessing takes place in the companion star (as a function of the orbital phase).

We have found two patterns of variability associated with short optical lags: single flares and `heartbeat-type' oscillations. Single short-lagged flares can be explained by accretion events producing an enhancement of the X-ray emission which is reprocessed at optical wavelengths within the binary system. `Heartbeat-type' oscillations were previously noted in the optical light curves of V404~Cyg during the period MJD~57199.840-57200.110 by \citet{kimura2016}.  Here we have also identified this pattern of variability in the system X-ray light curves and have studied when possible the correlation between them. During the MJD~57199.850--57199.905 epoch (see Fig.~\ref{fig:flick}, bottom right), we observed optical and X-ray `heartbeat-type' oscillations with periods of $\sim$6.4--8.1\,min. These oscillations are similar to the faster QPOs observed in X-rays for GRS~1915+105 and IGR~J17091--3624, with time-scales of tens of seconds to 1--2\, min \citep{belloni1997, neilsen2011, altamirano2011, janiuk2015}. These authors proposed that they can be caused by the Lightman-Eardley viscous instabilities in the inner accretion disc. At this epoch, \citet{jourdain2017} found that the \textit{INTEGRAL}/SPI spectra were compatible with a transition from a 'corona'-driven  to a 'jet'-driven regime. These disc instabilities could be related to the development of the jet. During this epoch, we measured short lags between the optical and X-ray emission ($\sim$1.2$\pm$0.3--1.5$\pm$0.3\,min for the soft and hard X-ray bands respectively). These optical delays are compatible with light travel times due to reprocessing in the accretion disc. Using the orbital ephemeris from \citet{casares1994} (P$_{orb}$ = 6.4714\,day and T$_{0}$ = 2448813.873 HJD), we can calculate the orbital phase  during this epoch, which is $\phi \sim$0.93, very close to superior conjunction, when the contribution from reprocessing in the companion star would be at a minimum. 

For those flares displaying optical lags longer than 2 minutes, the optical emission cannot be due to reprocessing of X-rays in the binary system. These long lags could be explained by X-ray reprocessing in the surrounding material (ejected as wind or jet) or by the interaction of blob ejections with this surrounding material or with previously ejected blobs. Re-heating of ejected blobs due to internal shocks has been observed in radio wavelengths in Sco~X--1 \citep{fomalont2001}, and re-brightening of ejected material has been seen in other microquasars, such as SS~433, in different energy ranges from X-rays \citep{migliari2005} to radio \citep{mioduszewski2004}.

Discrete jet ejections have been detected in V404~Cyg during the June 2015 outburst at radio and sub-mm wavelengths \citep{tetarenko2017} and inferred from optical and IR polarimetry \citep{shahbaz2016, tanaka2016}. \citet{tetarenko2017} found groups of 2--4 ejections (separated by at most $\sim$20\,min) with the ejecta travelling at bulk speeds varying between events (from 0.08 to 0.8\,$c$). Internal shocks between ejected blobs could produce some of the optical flares analysed in this work. Moreover, if the ejection is accompanied by an X-ray flare, positive lags would result due to the difference between the ejection time (X-ray flare), and the subsequent shock (optical flare). Re-brightening of ejected blobs by the irradiation by an intense X-ray flare could also lead to long-lagged optical flares.

The presence of surrounding material around V404~Cyg is known from previous studies of the June 2015 outburst, which suggest that a substantial fraction of the outer accretion disc was launched away by the strong disc winds detected at several epochs during the outburst in the optical \citep{munoz-darias2016} and in X-rays \citep{king2015}. These authors measured wind velocities of 1000--3000\,km/s, which would be compatible with the presence of material surrounding the system at distances varying from a few to tens of light--minutes over the outburst. The interaction of ejected blobs with the surrounding material or the re-brightening of this material by irradiation from an X-ray flare could produce also optical lags of several minutes. Potential reprocessing of X-rays in a source distant to the system, was inferred by \citet{walton2017} in the spectral fitting to \textit{NuSTAR} data. If reprocessing of X-rays takes place in a distant source, it would be expected to contribute to the optical emission also and some of the observed optical lags could be explained by this mechanism. Significant absorption by Compton-thick material was reported by \citet{motta2017,sanchez-fernandez2017,motta2017-a} at several epochs during the June 2015 outburst. X-ray reprocessing and extreme absorption values were also observed during the 1989 outburst \citep{oosterbroek1996,zycki1999}. The material ejected by the disc wind may be partly responsible for these extreme absorption values. It should be noted that occasionally an optical flare could be erroneously associated with an X-ray flare.  This could happen if the driving X-ray flare was obscured from our direct view, whereas the optical response would be visible as it arises in more distant regions that were not obscured.

It could happen also that some of the observed optical flares were also due to synchrotron emission from discrete ejections, as has been observed in the past from the correlations between the X-ray, near-IR and radio flares of GRS~1915+105 \citep{mirabel1998}. In this system, hard X-ray dips and soft X-ray flares on timescales of minutes were initially explained by repeated refilling and ejection of the inner accretion disc \citep{mirabel1998}. More recent studies have shown that it is the corona that is probably ejected during these events and the observed X-ray spectrum softening would mark the disappearance of the Comptonization component \citep{rodriguez2008-a, rodriguez2008}. A bright X-ray spike is usually detected during the recovery of the X-ray emission \citep{mirabel1998}, which coincide with the beginning of the IR flares. According to the synchrotron bubble model \citep{vanderlaan1966} the near-IR emission would originate as synchrotron emission from the ejections.

We have identified some multi-flare complex cases in which several optical and X-ray flares are observed, such as the interval MJD~57197.140--57197.240 (see Fig.~\ref{fig:X_opt_complex}, top). Just before flares 35--36, a polarized optical flare at MJD~57197.142 was measured by \citet{shahbaz2016}. These authors related this event with two radio flares observed at 16GHz and 5GHz peaking at $\sim$2\,h and $\sim$3.8\,h after the polarized flare respectively, and interpreted it as the birth of a major ejection event. According to our analysis, flares 35--36 display long lags with respect to the first two X-ray flares  ($\sim$34 and $\sim$26\,min; see Sect.~\ref{sub:X_opt}). Optical flares 35 and 36 might be related to jet ejection events, due either to synchrotron emission or to the interaction of these ejections with the surrounding material or with previous ejected blobs.
After flares 35--36, two simultaneous optical/X-ray flares (37--38) occur with very fast X-ray decay and slower optical decays. From the HR2 (defined between the 80--200\,keV and 20--80\,keV \textit{INTEGRAL}/IBIS energy bands), we can also see that flares 37--38 are harder and display faster decays than flares 35--36 (associated with the long lags), and display very bright relative fluxes in the 80--200\,keV band. Similar rapid non-thermal flares have also been observed in blazars and have been explained by Synchrotron Self Compton (SSC) processes in the jet \citep{sokolov2004}.

Another complex region identified during this outburst is the interval MJD~57198.200--57198.240, during which several optical and X-ray flares are observed (see Fig.~\ref{fig:X_opt_complex}, bottom). Optical flares 41 and 43 are simultaneous with two bright X-ray flares and flare 44 is simultaneous with a fainter X-ray flare. Optical flare 42 has a profile very similar to flare 41 and takes place during a period of faint X-ray emission. This flare displays a delay of $\sim$5\,min with respect to flare 41 and the simultaneous X-ray flare. Flare 41 corresponds to flare ``1'' in \citet{walton2017}. These authors proposed that the brightest flares observed in their \textit{NuSTAR} observations (including flare ``1'' and the subsequent X-ray flares) are produced by jet ejections. Furthermore, a reflection component from an unspecified distant region was also required in their X-ray spectral fitting. These authors also showed that the radio activity in the Arcminute Microkelvin Imager Large Array (AMI-LA) light curve starts just when the X-ray flaring activity begins in the \textit{NuSTAR} data. A state transition was also measured at this time by \citet{gandhi2017}, who found that subsecond optical/X-ray delays appeared together with a brightening of the radio jet. Flare 44 was explained by synchrotron 'cooling' processes by \citet{dallilar2017}. All these results suggest that various physical mechanisms, including jet ejections and reprocessing, could be contributing to the observed complex pattern.

\subsection{Optical emission preceding X-rays}
Finally, we have also observed intervals where a gradual increase of the optical emission and soft X-rays precede by several minutes a sudden rise of the hard X-ray flux  (marked with blue boxes in Figs.~\ref{fig:LCcomplete} and \ref{fig:LCcomplete2}, with zoomed-in views of some of them in Fig.~\ref{fig:opt_X}). We note that they usually occur after optical and X-ray low-flux states. Just at the end of these quiet intervals, the optical and soft X-ray fluxes brighten smoothly during several minutes before a sudden rise is detected in the hard X-ray light curve. As a consequence of this sudden rise, HR1 changes abruptly as well. One possible explanation to this previous soft state could be that an instability in the accretion flow approaches the BH producing an increase in the optical and soft X-ray emission and the sudden hardening could represent an ejection event. Previous identification of soft states were made by \citet{radhika2016} with Swift data for V404~Cyg at some points of the outburst, and these authors explained them as thermal disc emission. A complete spectral analysis, and a more detailed discussion of these events will be included in Kajava et al. (2018, in prep.).

Looking at the overall light curve evolution, we can see that these negative-lagged events usually take place before some long positive-lagged flares and before the symmetrical simultaneous double flares mentioned above (see Figs.~\ref{fig:LCcomplete} and \ref{fig:LCcomplete2}). In the picture described before, if the negative-lagged events are caused by an instability propagating through the accretion disk, the instability would eventually enter the corona and could result in discrete jet ejection events. Indeed, \citet{tetarenko2017} observed discrete ejections precisely just after one of the negative-lagged events we have identified at MJD\,57194.430. If discrete ejections are also produced after the other intervals when the optical and/or soft X-rays precede the hard X-rays (i.e. around MJD~57194.250, and after flares 18, 27, and 34), the presence of some single-lagged optical flares (e.g. flares 21, 22, 32, 33) and some double symmetric flares (e.g. flares 15--16, 19--20, 24--25, 28--29 and 30--31, and 35--36) that we observed subsequently could be explained by these discrete jet ejections. Some of these single and double optical flares occur simultaneously with the X-ray flares, while others appear delayed with respect to the X-ray flares. Different geometries and ejection angles could help to explain the differences between the time-delays found. The comparison with the radio emission during these epochs will be crucial for probing this scenario further.

\section{Summary and conclusions}
We have studied in detail the simultaneous optical and X-ray observations of the first V404~Cyg 2015 outburst from 18 June to 27 June 2015 performed with the IBIS, JEM-X, and OMC instruments onboard the \textit{INTEGRAL} observatory, complemented with additional data. From our analysis it is clear that the optical emission cannot always be explained by just one physical mechanism. We have classified the flares along the outburst with optical and X-ray observations according to the results of the cross-correlation analysis. We have distinguished between the following behaviours:

\begin{itemize}
 \item Simultaneous optical and X-ray emission. For most flares, we have measured time delays compatible with no lags between the optical and X-ray emission. We interpret this emission as coming from the same region or from close regions (less than one light--minute distances). Some of these flares display longer optical than X-ray flux decays, which could be produced by synchrotron emission and cooling from the corona or the base of the jet.
 \item Optical emission delayed from the X-rays: we found two different cases in which the optical emission appears to be delayed with respect to the X-rays:
 \begin{itemize}
 \item Short-lagged optical flares (with lags$<$2\,min). We have observed these lags in single flares and in intervals with flickering (`heartbeat-type') oscillations. We interpret these short lags as light travel times when optical emission is due to reprocessing of the X-ray emission in the accretion disc and/or in the companion star. In one of these flickering intervals (MJD~57199.850--57199.905), we have observed QPOs with frequencies of several minutes, and we have measured optical lags up to 1.5\,min. We propose that in this case, the optical lags are probably due to reprocessing of the X-ray emission on the external regions of the accretion disc. The mechanism producing these heartbeat-type oscillations could be related with disc instabilities and/or with the developing of a jet.
 \item Long-lagged optical flares (with lags from 3 to 55 minutes). We propose that some of these events are related to the interaction of discrete jet ejections with wind ejected material or with previously ejected blobs. In other cases, the optical flares could be produced by reprocessing of the X-rays in the surrounding material. Other flares could correspond to synchrotron emission from discrete jet ejections.
 \end{itemize}
 \item Optical emission preceding the X-ray emission. In some cases we have observed that the optical and soft X-rays brighten before a sudden brightening of the hard X-rays. Lags of $\sim$11\,min have been observed between both brightenings. These events take place before epochs in which there is radio evidence of jet ejections and/or we have observed optical flares delayed from the X-ray flares by several minutes. We propose this effect could be produced by some instability propagating in the accretion flow, which could eventually reach the corona and eventually lead to the subsequent observed jet ejections. 
\end{itemize}

From this analysis, we conclude that the optical emission observed for V404~Cyg during the June 2015 outburst is driven by a wide variety of mechanisms, including reprocessing of the X-ray emission into the accretion disc and the companion star, interaction of the ejected material with the surrounding material and/or with previously ejected blobs, and synchrotron emission from the jet. A more detailed multiwavelength analysis of each individual flare, ideally including radio observations and/or spectral information, should be performed to discriminate between the proposed scenarios.

\begin{acknowledgements}

The authors would like to thank the referee, Jan-Uwe Ness, for his very valuable comments and suggestions on the original manuscript. This work has been supported by Spanish MINECO grant ESP2015-65712-C5-1-R. This research has made use of data from the \textit{INTEGRAL}/OMC Archive at CAB (INTA--CSIC), pre-processed by ISDC. PAC is grateful to the Leverhulme Trust for the award of a Leverhulme Emeritus Fellowship. PG thanks STFC (ST/R000506/1) for support. We acknowledge with thanks the variable star observations from the AAVSO and VSNET International Databases contributed by observers worldwide and used in this research.

\end{acknowledgements}

\bibliographystyle{aa} % style aa.bst
\bibliography{allcites} % your references Yourfile.bib

\end{document}